\RequirePackage{fix-cm}
\documentclass[twocolumn]{svjour3}          % twocolumn
\smartqed  % flush right qed marks, e.g. at end of proof
\usepackage[colorlinks,linkcolor=blue,citecolor=blue,urlcolor=blue]{hyperref}
\usepackage{graphicx,epsfig,times}
\usepackage{amsmath,amsfonts,amssymb}
\usepackage{bbm}
\usepackage{subfigure,cases,multirow}
\usepackage{algpseudocode,algorithm,algorithmicx}
\usepackage[nocompress]{cite}

\journalname{Journal of Mathematical Imaging and Vision~~~}

\newcommand{\cB}{\mathcal B}
\newcommand{\cC}{\mathcal C}
\newcommand{\cU}{\mathcal U}
\newcommand{\cA}{\mathcal A}
\newcommand{\cM}{\mathcal M}
\newcommand{\cF}{\mathcal F}
\newcommand{\cL}{\mathcal L}
\newcommand{\cD}{\mathcal D}
\newcommand{\cV}{\mathcal V}
\newcommand{\cR}{\mathcal R}
\newcommand{\cT}{\mathcal T}
\newcommand{\cG}{\mathcal G}

\newcommand{\bR}{\mathbb R}
\newcommand{\bZ}{\mathbb Z}

\newcommand{\x}{\mathbf x}
\newcommand{\y}{\mathbf y}
\newcommand{\z}{\mathbf z}
\newcommand{\s}{\mathbf s}
\newcommand{\q}{\mathbf q}

\newcommand{\kc}{\mathfrak C}
\newcommand{\kp}{\mathbf p}
\newcommand{\kg}{\mathfrak g}
\newcommand{\ks}{\mathfrak s}
\newcommand{\kf}{\mathfrak F}

\newcommand{\fv}{\mathbf v}
\newcommand{\fu}{\mathbf u}
\DeclareMathOperator\interp{\mathbb I}

\begin{document}

\title{Fast Asymmetric Fronts Propagation for Image  Segmentation%\thanks{Grants or other notes
}
\titlerunning{Fast Asymmetric Fronts Propagation for Image  Segmentation}        % if too long for running head

\author{Da Chen  \and Laurent D. Cohen
}

\authorrunning{Da Chen \and Laurent D. Cohen} % if too long for running head

\institute{Da~Chen\and Laurent~D.~Cohen \at
              University Paris Dauphine, PSL Research University\\
              CEREMADE, CNRS, UMR 7534,  75016 Paris, France\\
               \email{chenda@ceremade.dauphine.fr}\\
               \email{cohen@ceremade.dauphine.fr}
}

\date{Received: 07-July-2017/ Accepted: 01-November-2017}

\maketitle
\begin{abstract}
In this paper, we introduce a generalized asymmetric fronts propagation model  based on the geodesic distance maps and the Eikonal partial differential  equations.  One of the key ingredients for  the computation of the geodesic distance map  is the geodesic metric, which can govern the action of the geodesic distance level set propagation. We consider a Finsler metric with the  Randers form, through  which the  asymmetry and anisotropy enhancements can be taken into account  to prevent the fronts leaking problem during the fronts propagation. These enhancements can be derived from the image edge-dependent vector field such as the gradient vector flow. 
The numerical implementations are carried out by the Finsler variant of the fast marching method, leading to very efficient interactive segmentation schemes. 
We apply the proposed Finsler fronts propagation model to image segmentation applications. Specifically,  the foreground and background segmentation is implemented by the Voronoi index map. In addition, for the application of tubularity segmentation, we exploit the level set lines of the geodesic distance map associated to the proposed Finsler metric  providing that a thresholding value is given.

\end{abstract}

\section{Introduction}
\label{sec:intro}
Fronts propagation models have been  considerably developed for the applications of image segmentation and boundary detection since the original level set framework proposed by Osher and Sethian \cite{osher1988fronts}. Guaranteed by  their  solid mathematical background, the fronts propagation models lead to strong abilities in a wide variety of computer vision tasks such as image segmentation and boundary detection~\cite{caselles1993geometric,malladi1995shape,caselles1997geodesic,yezzi1997geometric}.  In their basic formulation, the boundaries of an object  are modeled as closed contours, each of which can be  obtained   by evolving an initial closed curve in terms of a speed function till the stopping criteria reached.  A typical  speed function  usually involves a curve regularity penalty, for instance  the curvature, and  an image data term.  The use of  curve evolution scheme  for image segmentation can be backtrack to the original active contour model~\cite{kass1988snakes} which has inspired a amount of approaches~\cite{cohen1991active,xu1998snakes,cohen1997global,chen2017global,kimmel2003regularized,melonakos2008finsler}.

Let $\Omega\subset\bR^2$ be an open  bounded domain. Based on the level set framework~\cite{osher1988fronts}, a closed contour $\gamma$  can be retrieved by identifying the zero level set line of a scalar  function $\phi:\Omega\to\bR$ such that $\gamma:=\{\x\in\Omega;\,\phi(\x)=0\}$.  By this curve representation,  the curve evolution can be  carried out by evolving the function $\phi$ 
\begin{equation}
\label{eq:CurveEvolu}
\frac{\partial \phi}{\partial t}=\xi\,\|\nabla\phi\|,
\end{equation}
where $\xi:\Omega\to\bR$ is a speed function and $t$ denotes the time.  At any  time $t$, the curve $\gamma$ can be recovered by identifying the zero-level set line of the function $\phi$. Using the level set evolutional equation in Eq.~\eqref{eq:CurveEvolu},  the contours splitting and merging  can be adaptively handled. 
\begin{figure*}[t]
\centering
\includegraphics[width=17cm]{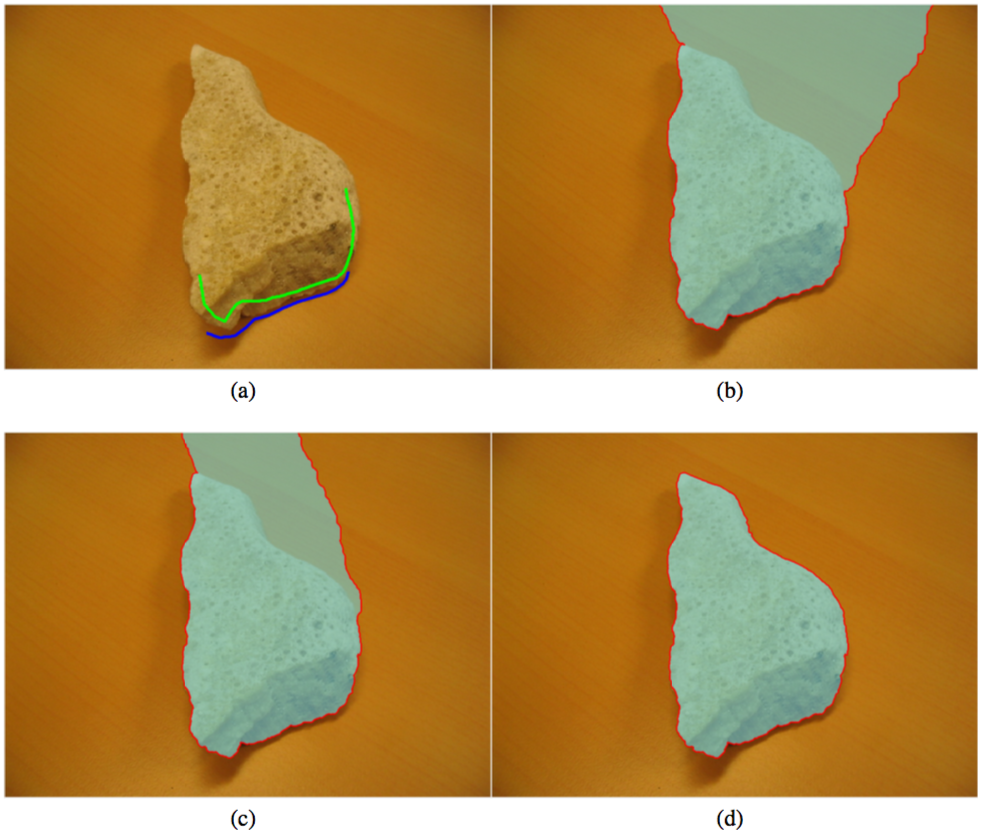}	
\caption{Fronts propagation  for interactive image segmentation  through different geodesic metrics. (\textbf{a}) The original image and the seeds,  where blue and green brushes indicate the seeds which  are placed in the  foreground and background,  respectively. (\textbf{b}) - (\textbf{d}) Segmentation results via the isotropic Riemannian metric, the anisotropic Riemannian metric and the  Finsler metric. The blue curves  represent the segmented foreground boundaries. }
\label{fig:Example}
\end{figure*}
The main drawback of  the level set-based front propagation method is its expensive computational burden. In order to alleviate this problem, Adalsteinsson and Sethian~\cite{adalsteinsson1995fast} suggested to  restrict the computation for the update of the level set function $\phi$ within a narrow band. In this case, only the values of $\phi$ at the points nearby the zero-level set lines are updated according to Eq.~\eqref{eq:CurveEvolu}. Moreover, the distance-preserving level set method~\cite{li2010distance} is able to avoid level set reinitialization  by enforcing the level set function $\phi$ as a signed Euclidean distance function from the current curves during the evolution. 

Despite the efforts devoted to the reduction in the computation burden,  the classical  level set-based fronts propagation scheme~\eqref{eq:CurveEvolu} is still impractical especially  for   real-time applications. In order to solve this issue, Malladi and Sethian~\cite{malladi1998real} proposed a new geodesic distance-based front propagation model for real-time image segmentation. 
It  relies on a geodesic distance map $\cU_\ks:\Omega\to\bR^+\cup\{0	\}$ associated to a set $\ks$ of source points. The value of $\cU_\ks(\x)$ at a point $\x$ in essence is equivalent  to the minimal geodesic curve length between the point $\x$ and a source point $\s\in\ks$ in the sense of an isotropic Riemannian metric, which is dependent on  a potential function $P:\Omega\to\bR^+$. The geodesic distance map $\cU_\ks$ coincides  with the  viscosity solution to the Eikonal equation, which can be efficiently computed by the fast marching methods~\cite{sethian1999fast,tsitsiklis1995efficient,mirebeau2014anisotropic,mirebeau2014efficient,mirebeau2017anisotropic}, leading  to a possible  real-time solution to the segmentation problem. In the context of segmentation, the potential function usually has  small values in the homogeneous  region and large values near the boundaries. Based on the geodesic distance map $\cU_\ks$,  a  curve can be denoted by  the $T$-level set  of  the distance map $\cU_\ks$, where $T>0$ is a geodesic distance thresholding value. In other words, a curve $\gamma$ can be characterized by the distance value $T$ such that  
\begin{equation}
\label{eq:CurveDisRep}
\gamma:=\{\x\in\Omega;\,\cU_\ks(\x)=T\}.
\end{equation}
By assigning large  values to the potential function $P$ around image edges, the basic idea behind~\cite{malladi1998real} is to use the curve $\gamma$ defined in Eq.~\eqref{eq:CurveDisRep}  to delineate the boundaries of interesting objects. One difficulty suffered by the geodesic distance-based fronts propagation scheme is that the fronts may leak outside  the targeted regions before  all the points of these regions have been visited by the fronts. The leakages sometimes occur  near the boundaries close to the source positions or in weak boundaries, especially when dealing with  long and thin structures. The main reason for this leaking  problem is the positivity constraint required by the metric (potential) functions for the  Eikonal equation. In order to solve this problem,  Cohen and Deschamps~\cite{cohen2007segmentation} suggested an adaptive freezing scheme for  tubular structure segmentation. They took into account a Euclidean curve length criterion to prevent  the fast marching fronts to travel outside the tubular structures in order  to avoid the leaking problem. The  main difficulty of this model lies at the choice of a suitable Euclidean curve length thresholding value. Chen and Cohen~\cite{chen2016vessel} considered an anisotropic Riemannian metric for vessel tree segmentation, where the vessel orientations are taken into account to mitigate the leaking problem.  Li and Yezzi~\cite{li2007local} proposed a dual fronts propagation model for active contours evolution, where the geodesic metric includes  both edge and region statistical information. The basic idea of~\cite{li2007local} is to propagate  the  fronts simultaneously from the  exterior and interior boundaries of the narrowband. The optimal contours can be recovered from the positions where the two fast marching fronts meet. These meeting interfaces also correspond to the boundaries of the adjacent Voronoi regions. 
Arbel\'aez and Cohen~\cite{arbelaez2004energy,arbelaez2008constrained} and Bai and Sapiro~\cite{bai2009geodesic} made use of  the  Voronoi index map  and  the Voronoi region for interactive image segmentation, both of which  can be constructed through the geodesic distance maps associated to   the pseudo path metrics. In their formulation, these models~\cite{arbelaez2004energy,arbelaez2008constrained,bai2009geodesic} allow the values of the metrics to be zero and to be dependent on path orientations.  The image segmentation can be characterized by the Voronoi regions, each of which involves all the points labeled by the same Voronoi index. In this case, the contours indicating the tagged object boundaries are no longer the level sets of the geodesic distance map, but the common boundaries of the adjacent Voronoi regions. 
Other interesting geodesic distance map-based image segmentation methods include~\cite{criminisi2008geos,price2010geodesic,cardinal2006intravascular}. 

A common point of the  Eikonal front propagation models mentioned above is that the segmentation procedure is carried out  by using   the  geodesic distance map itself. Finding geodesics  through the gradient descent on the geodesic distance map is an alternative way of using the Eikonal equation framework for practical applications. Since the original  work by Cohen and Kimmel~\cite{cohen1997global}, a broad variety of minimal path models have been proposed to solve various image analysis problems~\cite{benmansour2009fast,mille2015combination,chen2016finsler,li2007vessels,benmansour2011tubular,kaul2012detecting}. Recently, a significant contribution to the minimal path framework lies at the development of  the curvature-penalized geodesic models such as~\cite{bekkers2015pde,chen2017global,duits2016optimal,mashtakov2017tracking}. 
In the basic formulations of~\cite{duits2016optimal,chen2017global,chen2015global},  the curve length values  of the   minimal geodesics with curvature penalization  can be  approximately measured by  strongly anisotropic Riemannian metrics or Finsler metrics established in an orientation-lifted space.  
As a result, the geodesic distance maps associated to these metrics can be efficiently and accurately estimated by the Hamiltonian  fast marching method~\cite{mirebeau2017fast}.   The  curvature-penalized geodesics can be recovered via a gradient descent scheme on the associated geodesic distance map.

In this paper,   we extend the geodesic distance map-based front propagation framework  to a Finsler case, where both the edge anisotropy and asymmetry are taken into account simultaneously.  Our model thus relies on the geodesic distance map itself instead of minimal paths.  Moreover, we also present two ways to construct the Finsler metrics with respect to the applications of foreground and background object segmentation and tubularity segmentation. In order to quickly find suitable and reliable  solutions in various situations, it is  important for the fronts propagation models with a single-pass manner to be robust against to the  leaking problem.  The existing  front propagation approaches invoking either Riemannian metrics~\cite{cohen2007segmentation,chen2016vessel} or pseudo path metrics~\cite{arbelaez2004energy,arbelaez2008constrained,bai2009geodesic}, do not take into account  the edge asymmetry information. This may  increase the risk of the  leaking  problem especially when the provided seeds are close to the targeted boundaries. 
We show an example of the leaking problem in Fig.~\ref{fig:Example}. In  Fig.~\ref{fig:Example}a, the source points inside the  foreground and background are indicated by green and blue brushes, respectively. The  contours in Figs.~\ref{fig:Example}b and \ref{fig:Example}c obtained from the  Riemannian metrics  pass through the boundaries before the whole object has been covered by the fast marching fronts.  In contrast, the segmented contour derived from the proposed Finsler metric can  avoid such problem as shown in Fig.~\ref{fig:Example}d.

\subsection{Paper Outline}

The remaining of this paper is organized as follows: In Section~\ref{sec:Background}, we introduce the geodesic distance map  associated to a general Finsler metric, the  Voronoi   regions and the relevant numerical tool. Section~\ref{sec:basicFinslerMetric} presents the construction of the asymmetric Finsler metrics associated to different image segmentation applications.  The numerical considerations for the Finsler metrics-based fronts propagation are introduced in Section~\ref{sec:NumericalConsideration}. The experimental results and  the conclusion are respectively presented in Sections~\ref{sec:Experiment} and~\ref{sec:Conclusion}.

\section{Background on Geodesic Distance Map}
\label{sec:Background}

\begin{figure*}
\setcounter{subfigure}{0}
\centering
\includegraphics[width=17cm]{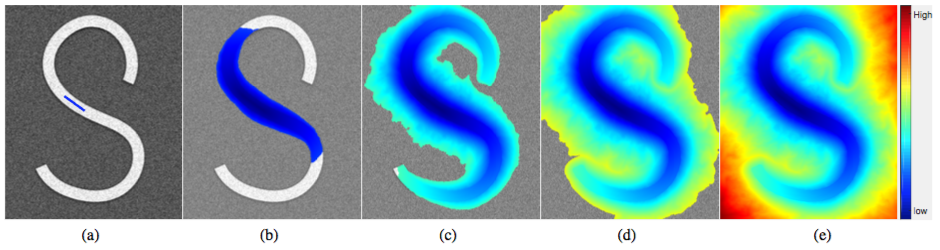}
\caption{The course of the fast marching front propagation for the computation of the geodesic distance map $\cU_\ks$. (\textbf{a}) The original image. The blue brush indicates the  source point set $\ks$  such that  $\cU_\ks(\x)=0,\,\forall\x\in\ks$. (\textbf{b})-(\textbf{e}) The course of the fast marching front propagation. }
\label{fig:FMPropagation}
\end{figure*}

A Finsler geodesic metric $\cF:\Omega\times\bR^2\to[0,+\infty]$ is a continuous function over the domain $\Omega\times\bR^2$.  For each fixed point $\x\in\Omega$, the geodesic metric $\cF(\x,\fv)$ can be  characterized by  an asymmetric norm of  $\fv\in\bR^2$. In other words, the Finsler geodesic  metric $\cF$ is convex and 1-homogeneous on its second argument. It is also  potentially asymmetric such  that   $\exists \x\in\Omega$ and $\exists \fv\in\bR^2$, the following inequality is held
\begin{equation}
\label{eq:AsymmetryDefinition}
\cF(\x,\fv)\neq\cF(\x,-\fv).
\end{equation}
The geodesic curve length associated to  the metric $\cF$ along  a Lipschitz continuous curve $\cC$  can be expressed by
\begin{equation}
\label{eq:Curvelength}
\ell_\cF(\cC):=	\int_\cC \cF(\cC(s),\cC^\prime(s))\,ds,
\end{equation}
where $\cC^\prime(s)=\frac{d}{ds}\cC(s)$ is the first-order derivative of the curve $\cC$ and  $s$ is the arc-length parameter of the curve $\cC$.

Letting $\ks\subset\Omega$ be a set which involves all the source points.  The minimal curve length  from $\y$ to $\x$ with respect to the Finsler metric $\cF$ is defined by
\begin{equation}
\label{eq:MCL}
\cD_{\cF}(\y,\x)=\inf_{\cC\in \cA_{\y,\x}} \ell_\cF(\cC),
\end{equation}
where $\cA_{\y,\x}$ is the set of all the  Lipschitz continuous curves linking from a  point $\y$ to $\x\in\Omega$. 

The geodesic distance map $\cU_\ks$ associated to  the geodesic metric $\cF$ can be defined in terms of the minimal  curve length $\cD_\cF$ in Eq.~\eqref{eq:MCL} such that 
\begin{equation}
\label{eq:MinimalActionMap}
\cU_\ks(\x):=\inf_{\y\in\ks}\,\cD_\cF(\y,\x).
\end{equation}
The geodesic distance map $\cU_{\ks}$ is  the  unique viscosity solution to the Eikonal equation~\cite{mirebeau2014efficient,chen2017global} 
\begin{equation}
\begin{cases}
\label{eq:HJEquation}
\displaystyle\max_{\|\fv\|\neq 0}\frac{\langle\nabla\cU_\ks(\x),\fv  \rangle}{\cF(\x,\fv)}=1,\, &\forall\x\in\Omega\backslash\ks,	\\
\cU_\ks(\x)=0,\,&\forall\x\in\ks,
\end{cases}
\end{equation}
where $\nabla\cU_\ks$ denotes the standard  Euclidean gradient of $\cU_\ks$ and $\langle\cdot,\cdot\rangle$ is the Euclidean scalar product in the Euclidean space $\bR^2$.

The Eikonal  equation~\eqref{eq:HJEquation} can be interpreted by the Bellman's optimality principle which states that 
\begin{equation}
\label{eq:BellmanOptimality}
\cU_\ks(\x)=\min_{\y\in\partial\Lambda(\x)}\,\{\cD_\cF(\y,\x)+\cU_\ks(\y)\},
\end{equation}
where $\Lambda(\x)\subset\Omega$ is a neighbourhood of point $\x$ and $\partial\Lambda(\x)$ is the boundary of $\Lambda(\x)$.  This interpretation is a key ingredient  for the numerical computation of the geodesic distance map via the fast marching method~\cite{mirebeau2014efficient}.

\subsection{Voronoi Index Map}
In this section, we consider a more general  case for which   a family of  source point sets are provided. Suppose that these source point sets are denoted by $\ks_k$ and are indexed by $k\in\{1,2,\cdots,n\}$ with $n$ the total number of source point sets. For the sake of simplicity, we note $\ks=\cup_{k=1}^n\ks_k$.

For a given geodesic metric $\cF$, we can compute the respective  geodesic distance map $\cU_k$ associated  to each source point set $\ks_k$ by Eq.~\eqref{eq:MinimalActionMap}.
A Voronoi index map is a function $\cL:\Omega\to\{1,2,\cdots,n\}$ such  that for any source point $\x\in\ks_k$
\begin{equation}
\cL(\x)=k,\quad k\in\{1,2,\cdots,n\},
\end{equation}
and for any domain point $\x\in\Omega\backslash\ks$, the Voronoi index $\cL(\x)$ is  identical to the index of the closest  source point set in the sense of the geodesic distance~\cite{arbelaez2004energy,benmansour2009fast}. One can   construct a Voronoi index map $\cL$ in terms of the geodesic distance maps $\cU_k$ ($1\leq k\leq n$) by 
\begin{equation}
\label{eq:VIM}
\cL(\x)=\underset{1\leq k \leq n}{\rm{arg\,min}}~~\cU_k(\x),\quad \forall\x\in\Omega.
\end{equation}
By the Voronoi index map $\cL$, the whole domain $\Omega$ can be partitioned  into $n$ Voronoi regions $\cV_k\subset \Omega$
\begin{equation}
\label{eq:VR}	
\cV_k:=\{\x\in\Omega;\,\cL(\x)=k\}.
\end{equation}
The common boundary $\Gamma_{i,j}:=\partial\cV_i\cap\partial\cV_j$ of two adjacent Voronoi regions $\cV_i$ and $\cV_j$ is comprised of  a collection of equidistant points to the source point sets $\ks_i$ and $\ks_j$, i.e., 
\begin{equation}
\cU_i(\x)=\cU_j(\x),\quad\forall \x\in\Gamma_{i,j}.
\end{equation}
Finally, we consider a  geodesic distance map $\cU_\ks$ associated to the set  $\ks=\cup_k\ks_k$ which can be expressed by
\begin{equation}
\label{eq:UniqueMinimalMap}
\cU_\ks(\x)=\min_{1\leq k\leq n}\,\cU_k(\x),	
\end{equation}
where $\cU_k$ is the geodesic distance map with respect to the source point set $\ks_k$ indexed by $k$.

\subsection{Fast Marching Method}
Many approaches~\cite{bornemann2006finite,weber2008parallel,yatziv2006n,rouy1992viscosity}  can be used to estimate the geodesic distance map $\cU_\ks$. Among them,  the fast marching method~\cite{sethian1999fast,tsitsiklis1995efficient}  solves the Eikonal equation in a very efficient way. It has a similar distance estimation scheme with  Dijkstra's graph-based shortest path algorithm~\cite{dijkstra1959note}.  One crucial  ingredient of the fast marching method is the construction of the stencil map $\Lambda$, where $\Lambda(\x)$  defines the neighbourhood  of a grid point $\x$. The isotropic fast marching methods~\cite{sethian1999fast,tsitsiklis1995efficient} are established  on an orthogonal  4-connectivity neighbourhood system, which  suffers some difficulties for the distance computation associated to asymmetric  Finsler metrics~\cite{mirebeau2014efficient}.  In order to find accurate solutions to the Finsler Eikonal equation, more complicated neighbourhood systems are taken into account~\cite{mirebeau2014anisotropic,mirebeau2014efficient,sethian2003ordered,mirebeau2017fast}. These neighbourhood systems or stencils are usually  constructed depending on the   geodesic metrics. In this paper,  we adopt the Finsler variant of the fast marching method proposed by Mirebeau~\cite{mirebeau2014efficient}. It invokes a geometry tool of  anisotropic stencil refinement and leads to  a highly accurate solution, but requires a low computation complexity.

\subsubsection{Hopf-Lax Operator for Local Distance Update}

The Finsler variant of the fast marching method~\cite{mirebeau2014efficient} estimates the geodesic distance map  on  a regular discretization grid $\bZ^2$ of the  domain $\Omega$. It makes use of the Hopf-Lax operator to approximate~\eqref{eq:BellmanOptimality} by
\begin{equation}
\label{eq:HopfLax}
\cU_\ks(\x)=\min_{\y\in\partial \Lambda(\x)}\{\cF(\x,\x-\y)+\interp_{\Lambda(\x)}\cU_\ks(\y)\},
\end{equation}
where  $\Lambda(\x)$ denotes the stencil of $\x$ involving a set of vertices in $\bZ^2$ and   $\interp_{\Lambda(\x)}$ is a piecewise linear interpolation operator in the neighbourhood $\Lambda(\x)$. The estimate of the quality and order for the solution to~\eqref{eq:HopfLax} can be found in~\cite{mirebeau2014efficient}.

The Hopf-Lax operator is first introduced for the geodesic distance computation by Tsitsiklis~\cite{tsitsiklis1995efficient} from an optimal control point of view.  The minimal curve length $\cD_\cF$ of a short geodesic from $\y$ to $\x$ is approximated by the length $\cF(\x,\x-\y)$ of a line segment $\overline{\x\y}$. The geodesic distance value $\cU_\ks(\y)$ in Eq.~\eqref{eq:BellmanOptimality} is estimated by a piecewise linear interpolation operator $\interp_{\Lambda(\x)}$ at  $\y$ located at  the stencil boundary $\partial \Lambda(\x)$. It  is comprised of a set $\cT_\x$ of one-dimensional simplexes or line segments. Each simplex $\mathbb{T}_i\in\cT_\x$ connects two adjacent vertices  which are involved in the stencil  $\Lambda(\x)$.
The solution $\cU_\ks$ to the Hopf-Lax operator~\eqref{eq:HopfLax} can be attained  by
\begin{equation}
\label{eq:HopfLax2}
\cU_\ks(\x)=\min_{\mathbb{T}_i\in\cT_\x}\,U_i(\x),
\end{equation}
 where $U_i$ is the solution to the  minimization problem
\begin{equation}
\label{eq:HopfLax3}
U_i(\x)	=\min_{\y\in \mathbb{T}_i}\{\cF(\x,\x-\y)+\interp_{\Lambda(\x)}\cU_\ks(\y)\}.
\end{equation}
For  each simplex $\mathbb{T}_i\in\cT_\x$ which joins two vertices $\z_1$ and $\z_2$,  the minimization problem~\eqref{eq:HopfLax3} can be approximated by Tsitsiklis' theorem~\cite{tsitsiklis1995efficient} such that 
\begin{equation}
\label{eq:TsitsiklisFormula}
U_i(\x)=\min_{\vec\lambda}\,\cF\left(\x,\x-\sum_{i=1}^2\lambda_i\z_i\right)+\sum_{i=1}^2\lambda_i\cU_{\ks}(\z_i),	
\end{equation}
 where $\vec\lambda=(\lambda_1,\lambda_2)$ subject to $\lambda_1,\lambda_2\geq 0$ and $\sum^2_i\lambda_i=1$.

\subsubsection{Fast Marching Fronts Propagation Scheme}
%----algorithm------%
\begin{algorithm}[t]
\caption{Fast Marching Fronts Propagation}
\label{alg:VIM}
\begin{algorithmic}[1]
\renewcommand{\algorithmicrequire}{ \textbf{Input:}}
\renewcommand{\algorithmicensure }{ \textbf{Output:}}
\Require Source points set $\ks=\cup_k\ks_k$.
\Ensure  Geodesic distance map $\cU_\ks$ and Voronoi index map $\cL$.
\State $\forall \x\in\Omega\backslash\ks$,  set $\cU_\ks(\x)\gets\infty$ and $b(\x)\gets$\emph{Far}.       
\State $\forall \x\in\ks$,  set $\cU_\ks(\x)\gets 0$ and $b(\x)\gets$\emph{Trial}.
\State $\forall \x\in\ks_k$, set $\cL(\x)=k$.
\While{there remains at least one \emph{Trial}  point}
\State Find a \emph{Trial} point  $\x_{\rm min}$ globally minimizing $\cU_\ks$.
\State Set $b(\x_{\rm min})\gets$\,\emph{Accepted}.
\If{$\x_{\rm min}\notin\ks$} 
\State Update the Voronoi index $\cL(\x_{\rm min})$ by Eq.~\eqref{eq:NumVIM}.
\EndIf
\For{all $\z\in\bZ^2$ such that $\x_{\rm min}\in\Lambda(\z)$}
\label{line:MetricUpdate}
\If{$b(\z)\neq$\emph{Accepted} and $\z\notin\ks$}
\State $/*$ \emph{Update some map} $\kc_{\rm dyn}(\z)$ \emph{if necessary}. $*/$    
\label{line:DynamicUpdate}   
\State Find $\hat\cU(\z)$ by evaluating the Hopf-Lax formula~\eqref{eq:HopfLax2}.
\State Set $\cU_\ks(\z)\gets\min\{\cU_\ks(\z),\hat\cU(\z)\}$ and $b(\z)\gets$\,\emph{Trial}.
\EndIf
\EndFor
\EndWhile 
\end{algorithmic}
\end{algorithm}

The fast marching method estimates the geodesic distance map $\cU_{\ks}$ in a wave front propagation manner. We demonstrate the course of the fast marching fronts  propagation in Fig~\ref{fig:FMPropagation} on a synthetic image. In this figure, we invoke a Finsler metric for the computation of the geodesic distance map $\cU_\ks$, where the metric will be presented in Section.~\ref{sec:NumericalConsideration}.
The fast marching fronts propagation is coupled with a procedure of label assignment operation, during which  all the grid points are classified into three categories:
\begin{itemize}
	\item \emph{Accepted} points,  for which the values of $\cU_\ks$ have been estimated and frozen.
\item \emph{Far} points, for which the values of $\cU_\ks$ are unknown.
\item \emph{Trial} points, which are the remaining grid points in $\bZ^2$ and  form  the fast marching \emph{fronts}. A \emph{Trial} point will be assigned a label  of \emph{Accepted} if it has the minimal geodesic distance value among all the \emph{Trial} points.
\end{itemize}
In the course of the geodesic distance estimation,  each grid point $\x\in\bZ^2\backslash\ks$ will be visited by the monotonically  advancing  fronts  which expand from the source  points involved in $\ks$. 
The  values of $\cU_\ks$ for all the \emph{Trial}  points are stored in a priority queue in order to quickly find the point with minimal $\cU_\ks$. The label assignment procedure\footnote{Initially, each source point  $\x\in\ks$ is tagged as \emph{Trial} and  the remaining grid points are tagged as \emph{Far}. }  can be carried out by a binary map $b:\bZ^2\to\{\emph{Accepted},\,\emph{Far},\,\emph{Trial}\}$.    

Suppose that  $\ks=\cup_k\ks_k$ with $\ks_k$ a source point set. The geodesic distance map $\cU_\ks$ and the Voronoi index map $\cL$ can be simultaneously computed~\cite{benmansour2009fast,cohen2001multiple}, where the  computation scheme in each iteration can be divided into two steps.  

\noindent{\textbf{Voronoi index update}}.
In each geodesic distance  update iteration, among all the \emph{Trial} points, a  point  $\x_{\rm min}$ that globally minimizes the geodesic distance map $\cU_\ks$ is chosen and tagged as \emph{Accepted}. We set $\cL(\x_{\rm min})=k$  if $\x_{\rm min}\in\ks_k$. Otherwise, the geodesic distance value $\cU_\ks(\x_{\rm min})$ can be estimated in the simplex $\mathbb T^*\in\cT_{\x_{\rm min }}$ (see Eq.~\eqref{eq:HopfLax2}), where the vertices relevant to $\mathbb T^*$ are respective $\z_1$ and $\z_2$. This is done by finding the  solution to~\eqref{eq:TsitsiklisFormula} with respect to the simplex $\mathbb T^*$, where the corresponding minimizer is  $\vec\lambda^*=(\lambda^*_1,\lambda^*_2)$. Then the Voronoi index map $\cL$ can be computed by
 \begin{equation}
\label{eq:NumVIM}
\cL(\x_{\rm min})=
\begin{cases}
\cL(\z_1),\quad  &\text{if}~\lambda_1^*\geq\lambda^*_2,\\
\cL(\z_2),\quad &\text{otherwise}.
\end{cases}	
\end{equation}

\noindent{\textbf{Local Geodesic Distance Update}}.
For a grid point $\x$, we denote by $\Lambda_\star(\x):=\{\z\in\bZ^2;\x\in\Lambda(\z)\}$ the reverse stencil.
The remaining step in this iteration is to update $\cU_{\ks}(\z)$ for each grid point $\z$ such that $\z\in\Lambda_\star(\x_{\rm min})$ and $b(\z)\neq$\emph{Accepted} through  the solution   $\hat\cU_\ks(\z)$ to the Hopf-Lax operator~\eqref{eq:HopfLax}. This is done by assigning  to $\cU_\ks(\z)$ the smaller value  between the solution $\hat\cU_\ks(\z)$ and the current geodesic distance value of $\cU_\ks(\z)$. Note that the solution $\hat\cU_{\ks}(\z)$ to~\eqref{eq:HopfLax} is attained using the stencil $\Lambda(\z)$~\cite{mirebeau2014efficient}.

The algorithm for the fast marching method is described in Algorithm~\ref{alg:VIM}. In this algorithm,  the computation of a map $\kc_{\rm dyn}$ in Line~\ref{line:DynamicUpdate} of Algorithm~\ref{alg:VIM} is not necessary for the general fast marching fronts propagation scheme, but required by our method as discussed in Section~\ref{subsec:UpdatePotential}.

\noindent{\textbf{Computation Complexity}}.
The complexity estimate for the isotropic fast marching methods~\cite{sethian1999fast,tsitsiklis1995efficient} established on the 4-neighbourhood system is bounded by $\mathcal O(N\ln N)$, where $N$ denotes the cardinality $N:=\#\bZ^2$ of the discrete domain $\bZ^2\cap \Omega$.  The complexity estimates for the Finsler variant cases~\cite{sethian2003ordered,mirebeau2014efficient} with adaptive stencil system  rely on the anisotropic ratio $\kappa(\cF)$ of the geodesic metric $\cF$. The estimate for the ordered upwind method~\cite{sethian2003ordered} is bounded by $\mathcal O(\kappa(\cF) N\ln N)$, which is impractical for image segmentation application.
In contrast,  for the method~\cite{mirebeau2014efficient} used in this paper, the complexity  bound reduces to $\mathcal O(N\ln\kappa(\cF)+N\ln N)$.
Note that the anisotropic  ratio  $\kappa(\cF)$ is defined by 
\begin{equation}
\label{eq:AnisoRatio}
\kappa(\cF):=\max_{\x\in\bZ^2}\left\{\max_{\|\fu\|=\|\fv\|=1}\frac{\cF(\x,\fu)}{\cF(\x,\fv)}	\right\}.
\end{equation}

%
%
%The C++ codes for the adaptive stencils-based fast marching methods~\cite{mirebeau2014anisotropic,mirebeau2014efficient,mirebeau2017anisotropic} can be downloaded from \href{https://github.com/Mirebeau}{Github-AnisoFM}.

\section{Finsler Metric Construction}
\label{sec:basicFinslerMetric}
\begin{definition}
Let  $S^+_2$ be  the collection of all the positive definite symmetric matrices with size $2\times 2$. For any matrix $M\in S_2^+$, we define a norm $\|\fu\|_M=\sqrt{\langle\fu,\,M\fu\rangle},\,\forall\fu\in\bR^2$.
\end{definition}

\subsection{Principles for Finsler Metric Construction}
In this section, we present  the construction method of the Finsler metric which is suitable for fronts propagation and image segmentation.  Suppose that  a vector field $\kg:\Omega\to\bR^2$ has been provided such that $\kg(\x)$ points to  the object edges at least when  $\x$ is nearby  them. In this case, the orthogonal vector field $\kg^\perp$ indicates  the tangents of the edges. 

Basically, the Eikonal equation-based fronts propagation models~\cite{malladi1998real} perform the segmentation scheme through a geodesic distance map. In order to find a good solution for image segmentation,  the used  geodesic metric should be able to reduce the risk of front leaking problem. For this purpose, we search for a direction-dependent  metric  $\cF_\kg$ satisfying the following inequality 
\begin{equation}
\label{eq:Criterion}
\cF_\kg(\x,\kg^\perp(\x))<\cF_\kg(\x,\kg(\x)) < \cF_\kg(\x,-\kg(\x)).
\end{equation}  
Recall that for an edge point $\x$, both the feature vectors $\kg^\perp(\x)$ or $-\kg^\perp(\x)$ are propositional to the tangent of the edge at $\x$. When the fast marching front arrives at the vicinity of image edges, it prefers to travel along the edge feature vectors $\kg^\perp(\x)$ and $-\kg^\perp(\x)$, instead of passing through the edges, i.e., prefers to travel along the direction $-\kg(\x)$.

The  inequality~\eqref{eq:Criterion} requires  the geodesic metric $\cF_\kg$ to be anisotropic and asymmetric with respect to its second argument. Thus, we consider a Finsler metric with a Randers  form~\cite{randers1941asymmetrical} involving a symmetric  quadratic term and a linear asymmetric term for any $\x\in\bR^2$ and any vector $\vec u\in\bR^2$
\begin{equation}
\label{eq:RandersMetrics}
\cF(\x,\fu):=\kc(\x)\left(\|\fu\|_{\cM_\kg(\x)}-\langle\vec\omega_\kg(\x),\fu\rangle\right),
\end{equation}
 where $\cM_\kg:\Omega\to S^+_2$ is a positive symmetric definite tensor field and  $\vec\omega_\kg:\Omega\to\bR^2$ is a vector field that is sufficiently small.   The function $\kc:\Omega\to\bR^+$ is a positive scalar-valued potential  which gets small values in the homogeneous regions and large values around the image  edges. It  can be derived from  the image data such as the coherence measurements of the image features, which will be discussed in detail in Section~\ref{subsec:UpdatePotential}. 

The tensor  field $\cM_\kg$ and the vector field  $\vec\omega_\kg$ should satisfy  the constraint 
\begin{equation}
\label{eq:PositiveConstraint}
\|\vec\omega_\kg(\x)\|_{\cM_\kg^{-1}(\x)}<1,\,\forall \x\in\Omega,	
\end{equation}
in order to guarantee the positiveness~\cite{mirebeau2014efficient} of the Randers  metric $\cF$.

\begin{figure*}[t]
\setcounter{subfigure}{0}
\centering
\includegraphics[height=7cm]{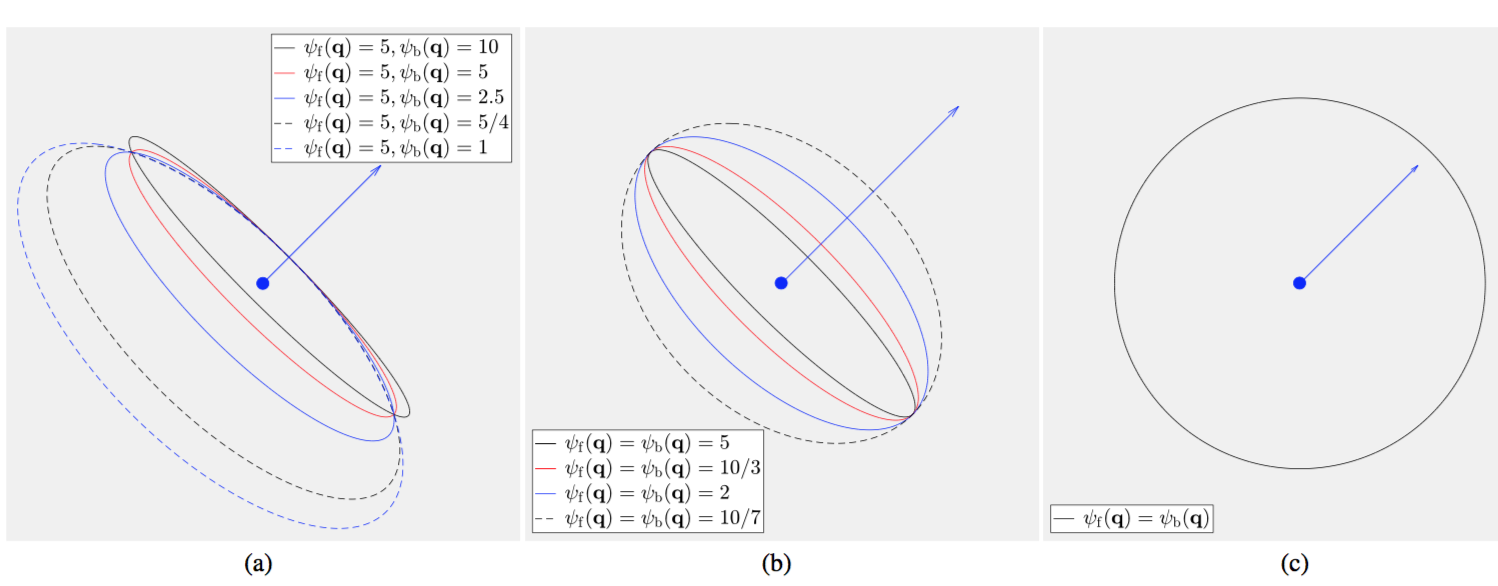}
\caption{Control sets $\cB(\q)$ for different  geodesic metrics associated to different values of $\psi_{\rm f}(\q)$ and $\psi_{\rm b}(\q)$. The blue arrow indicates the direction $\kg(\q)=(\cos(\pi/4),\sin(\pi/4))^{T}$.  The blue dots and the contours denote the origins and the boundaries of these control sets, respectively.  (\textbf{a}) Control sets for  Randers metrics associated to  $\psi_{\rm f}(\q)=5$ and different values of $\psi_{\rm b}(\q)$. (\textbf{b})  Control sets for  anisotropic Riemannian metrics with $\psi_{\rm f}(\q)=\psi_{\rm b}(\q)>1$. (\textbf{c}) Control set for an isotropic Riemannian metric with $\psi_{\rm f}(\q)=\psi_{\rm b}(\q)=1$.
}
\label{fig:MetricBalls}	
\end{figure*}   

We reformulate  the Randers metric $\cF_\kg$ in Eq.~\eqref{eq:RandersMetrics} as
\begin{equation}
\label{eq:FinslerComponents}
\cF_\kg(\x,\fu)	=\kc(\x)\,\cG_\kg(\x,\fu),
\end{equation}
where $\cG_\kg:\Omega\times\bR^2\to[0,\infty]$ is still  a Randers metric which can be formulated by
\begin{equation}
\label{eq:StaticRandersMetric}
\cG_\kg(\x,\fu)=\|\fu\|_{\cM_\kg(\x)}-\langle\vec\omega_\kg(\x),\fu\rangle.
\end{equation}
The remaining part of this section will be devoted to the construction of the Randers metric $\cG_\kg$ in terms of the vector field $\kg$ which is able to characterize the  directions  orthogonal to the image edges.

Let us define a new vector field $\bar\kg:\Omega\to\bR^2$ by
\begin{equation*}
\bar\kg(\x):=\frac{\kg(\x)}{\|\kg(\x)\|^2}.
\end{equation*}
The tensor field $\cM_\kg$ used  in Eq.~\eqref{eq:RandersMetrics} can be constructed dependently  on two scalar-valued coefficient functions $\eta_1$ and $\eta_2$ such that
\begin{equation}
\label{eq:Construction}
\cM_\kg(\x)=\eta_1^2(\x)\bar\kg(\x)\otimes\bar\kg(\x)+\eta_2(\x)\bar{\kg}^\perp(\x)\otimes\bar{\kg}^\perp(\x),
\end{equation}
where $\bar\kg^\perp(\x)$ is the orthogonal  vector of $\bar\kg(\x)$ and  $\otimes$ denotes the tensor product, i.e., $\fu\otimes\fu=\fu\fu^T$. Note that the eigenvalues of $\cM_\kg(\x)$ are $\eta_1^2(\x)/\|\kg(\x)\|^2$ and $\eta_2(\x)/\|\kg(\x)\|^2$, respectively corresponding to the eigenvectors $\kg(\x)/\|\kg(\x)\|$ and $\kg^\perp(\x)/\|\kg(\x)\|$. 

The  vector  $\vec\omega_\kg(\x)$ is positively collinear  to field $\kg(\x)$ for all $\x\in\Omega$ 
\begin{equation}
\label{eq:Construction2}
\vec\omega_\kg(\x)=\tau(\x)\,\bar\kg(\x),
\end{equation}
where $\tau:\Omega\to\bR$ is a scalar-valued coefficient function.

\begin{figure*}[t]
\setcounter{subfigure}{0}
\centering
\includegraphics[height=7cm]{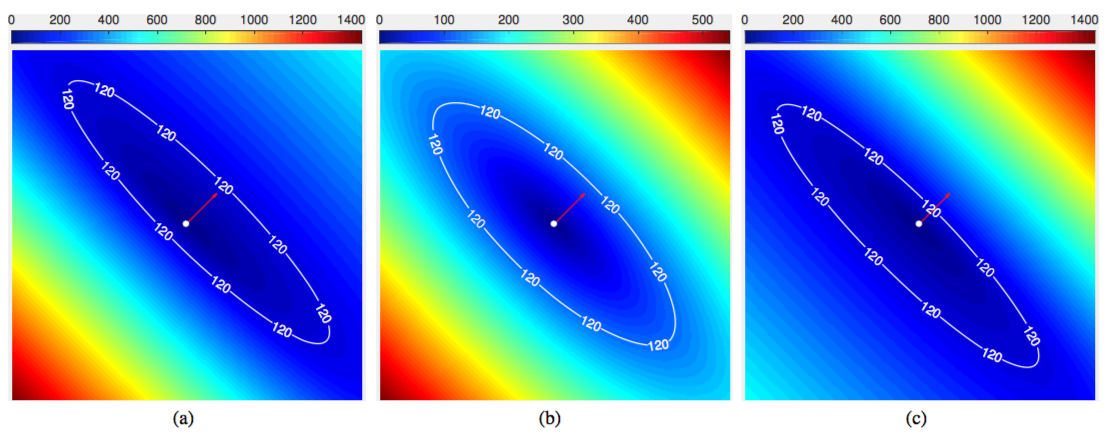}
\caption{Geodesic distance maps associated to the Randers metric $\cG_\kg$ with different values of $\psi_{\rm f}$ and $\psi_{\rm b}$. The red arrow indicate the vector $(\cos(\pi/4),\sin(\pi/4))^T$.  The white dots are the source points. Each white curve indicates a level set line of the respective geodesic distance map.  (\textbf{a}) Geodesic distance map associated to $\psi_{\rm f}\equiv3$ and $\psi_{\rm b}\equiv 8$. (\textbf{b}) Geodesic distance map associated to $\psi_{\rm f}\equiv 3$ and $\psi_{\rm b}\equiv 3$.  (\textbf{c}) Geodesic distance map associated to $\psi_{\rm f}\equiv 8$ and $\psi_{\rm b}\equiv3$. The color bars are on the top of each figure.}
\label{fig:maps}
\end{figure*}

We estimate the  coefficient functions $\eta_1$, $\eta_2$ and $\tau$ through  two  cost  functions $\psi_{\rm f},\,\psi_{\rm b}:\Omega\to (1,\,+\infty)$, which  assign the cost values $\psi_{\rm f}(\x)$, $\psi_{\rm b}(\x)$ and $1$ to the Randers metric $\cG_\kg$ respectively along the directions $\kg(\x)$,  $-\kg(\x)$ and $\kg^\perp(\x)$ for any point $\x\in\Omega$ such that
\begin{equation}
\label{eq:FinslerianConstruction}
\begin{cases}
\cG_\kg(\x,\kg(\x))&= \psi_{\rm f}(\x),\\
\cG_\kg(\x,-\kg(\x))&=\psi_{\rm b}(\x),\\
\cG_\kg(\x,\kg^\perp(\x))&=1.		
\end{cases}
\end{equation}
Combining Eqs.~\eqref{eq:Construction} and \eqref{eq:FinslerianConstruction} yields that 
\begin{equation}
\label{eq:Coefficient}
\begin{cases}
\eta_1(\x)-\tau(\x)&= \psi_{\rm f}(\x),\\
\eta_1(\x)+\tau(\x)&=\psi_{\rm b}(\x),\\
\eta_2(\x)&= 1,	
\end{cases}
\end{equation}
for any $\x\in\Omega$. 

The positive symmetric definite  tensor field  $\cM_\kg$ and the vector field $\vec\omega_\kg$  thus can be respectively expressed  in terms of the cost functions $\psi_{\rm f}$ and $\psi_{\rm b}$ by 
\begin{align}
\label{eq:RewrittenTensor}
\cM_\kg(\x)=&\frac{1}{4}(\psi_{\rm f}(\x)+\psi_{\rm b}(\x))^2\,\bar\kg(\x)\otimes\bar\kg(\x)\nonumber\\
&+\bar\kg^\perp(\x)\otimes\bar\kg^\perp(\x),
\end{align}
and 
\begin{equation}
\label{eq:RewrittenVectorFiled}	
\vec\omega_\kg(\x)=\frac{1}{2}(\psi_{\rm b}(\x)-\psi_{\rm f}(\x))\,\bar\kg(\x).	
\end{equation}
Based on the tensor field $\cM_\kg$ and the vector field $\vec\omega_\kg$ respectively formulated in Eqs.~\eqref{eq:RewrittenTensor} and \eqref{eq:RewrittenVectorFiled},  the positiveness constraint~\eqref{eq:PositiveConstraint}  is satisfied due to  the assumption that $\psi_{\rm f}(\x)>1$ and $\psi_{\rm b}(\x)>1,\,\forall \x\in\Omega$. The cost functions $\psi_{\rm f}$ and $\psi_{\rm b}$  can be derived from the image edge information such as the image gradients, which will be discussed in Section~\ref{sec:NumericalConsideration}. 

Note that if we set $\psi_{\rm f}\equiv\psi_{\rm b}$, the vector field $\vec \omega_\kg$ will vanish, i.e., $\vec\omega_\kg\equiv\mathbf 0$ (see Eq.~\eqref{eq:RewrittenVectorFiled}). In this case, one has $\langle\vec\omega_\kg(\x),\fu \rangle= 0$ for any point $\x\in\Omega$ and any vector $\fu\in\bR^2$, leading to  a special form of the Randers metric $\cG_\kg$. This special form is a symmetric (potentially anisotropic)  Riemannian metric $\cR(\x,\fu)=\|\fu\|_{\cM_{\kg}(\x)}$ which depends only on the tensor field $\cM_{\kg}$.

\begin{remark}[\textbf{Randers Eikonal Equation}]
The Eikonal equation for a general Finsler metric can be found in Eq.~\eqref{eq:HJEquation}.  Associated  to the Randers metric $\cG_{\kg}$~defined in~\eqref{eq:StaticRandersMetric}, the general Eikonal equation~\eqref{eq:HJEquation} gets to be  the following form 
\begin{equation}
\label{eq:RandersEikonal}
\begin{cases}
\|\nabla\cU_\ks(\x)-\vec\omega_\kg(\x)\|_{\cM^{-1}_{\kg}(\x)}=1,\quad&\forall\x\in\Omega\backslash\ks,\\
\cU_\ks(\x)=0,\quad&\forall \x\in\ks,
\end{cases}
\end{equation}
where $\cU_\ks$ is the geodesic distance map and $\ks$ is the source point set.
\end{remark}

\subsection{Tissot's indicatrix}
A basic tool for studying and visualizing the geometry distortion induced from a geodesic metric is the Tissot's indicatrix defined as the collection of control sets in the tangent space~\cite{chen2017global}. For an arbitrary  geodesic metric $\cF:\Omega\times\bR^2\to[0,\infty]$, the  control set $\cB(\x)$ for any point $\x\in\Omega$ is defined as the unit ball centered at $\x$ such that
\begin{equation}
\label{eq:UnitBall}
\cB(\x):=\{\fu\in\bR^2;\cF(\x,\fu)\leq 1\}.	
\end{equation}
We demonstrate the  control sets  $\cB(\q)$ in Fig.~\ref{fig:MetricBalls} for the Randers metric $\cG_\kg(\q,\cdot)$ with different values of  $\psi_{\rm f}(\q)$ and $\psi_{\rm b}(\q)$ at a point $\q\in\Omega$.  The vector  $\kg(\q)$ is set as 
\begin{equation*}
\kg(\q)=\left(\cos\left(\frac{\pi}{4}\right),\sin\left(\frac{\pi}{4}\right)\right)^{T}.	
\end{equation*}
In Fig.~\ref{fig:MetricBalls}a,  we show the control sets for the  Randers metric $\cG_\kg$ with respect to  different  values of $\psi_{\rm b}(\q)$ and a fixed value  $\psi_{\rm f}(\q)=5$. One can point out that the common origin of these control sets have shifted from the original  center of the ellipses\footnote{These ellipses  are the boundaries of the control sets.}  due to the asymmetric property as formulated in Eq.~\eqref{eq:AsymmetryDefinition}. In Fig.~\ref{fig:MetricBalls}b, the control sets for the Randers metric $\cG_\kg$ associated to   $\psi_{\rm f}(\q)=\psi_{\rm b}(\q)>1$ are demonstrated, where $\cG_\kg(\q,\cdot)$ gets to be anisotropic and symmetric on its second argument. When $\psi_{\rm f}(\q)=\psi_{\rm b}(\q)=1$, the values of the Randers metric $\cG_\kg(\q,\vec u)$ turn to be invariant  with respect to $\vec u$ as shown  in Fig.~\ref{fig:MetricBalls}c. In this case,  the tensor $\cM_\kg(\q)$ is propositional to the identity matrix. 

In Fig.~\ref{fig:maps}, we show the geodesic distance maps associated to $\cG_\kg$ with different values of the cost functions $\psi_{\rm f}$ and $\psi_{\rm b}$. The vector field $\kg$ is set to 
\begin{equation*}
\kg\equiv\left(\cos\left(\frac{\pi}{4}\right),\sin\left(\frac{\pi}{4}\right)\right)^T.
\end{equation*}
In Figs.~\ref{fig:maps}a and ~\ref{fig:maps}c,  we can see that the geodesic distance maps have strongly asymmetric and anisotropic appearance. In Fig.~\ref{fig:maps}b, we observe that the geodesic distance  map appears to be symmetric and strongly anisotropic. This is because the respective propagation  speed of the  fast marching fronts  along the directions $(\cos(\pi/4),\sin(\pi/4))^T$ and $-(\cos(\pi/4),\sin(\pi/4))^T$ is identical to each other.

\section{Data-Driven Randers Metrics Construction}
\label{sec:NumericalConsideration}

\subsection{Cost Functions $\psi_{\rm f}$ and $\psi_{\rm b}$ for the Application of Foreground and Background Segmentation}
\label{subsec:ObjectData}
In this section, we present the computation method for  the  cost functions $\psi_{\rm f}$ and  $\psi_{\rm b}$ using the image edge information, based on which the image data-driven Randers metric $\cG_{\kg}$ can be obtained.

Let $\mathbf I=(I_1,I_2,I_3):\Omega\to\bR^3$ be a vector-valued image in the chosen color space and let $G_\sigma$ be a Gaussian kernel with variance $\sigma$ (We set  $\sigma=1$ through all the experiments of this paper). The gradient  of the image $\mathbf I$ at each  point $\x=(x,y)$ is a $2\times 3$ Jacobian matrix $$\nabla\mathbf I_\sigma(\x)=\nabla G_\sigma\ast \mathbf I\,(\x)$$ involving the Gaussian-smoothed first-order derivatives along the  axis directions $x$ and $y$
\begin{equation}
\label{eq:FNorm}
\nabla\mathbf I_\sigma(\x)=
\begin{pmatrix}
\partial_x G_\sigma\ast I_1~~&\partial_x G_\sigma\ast I_2~~&\partial_x G_\sigma\ast I_3\\
\partial_y G_\sigma\ast I_1~~&\partial_y G_\sigma\ast I_2~~&\partial_y G_\sigma\ast I_3\\
\end{pmatrix}(\x).
\end{equation}
Let $\rho:\Omega\to\bR$ be an edge saliency  map. It has high values in the vicinity of image edges and low values inside the flatten regions.  For each domain point $\x\in\Omega$, the value of $\rho(\x)$ can be computed by  the Frobenius norm of the Jacobian matrix $\nabla\mathbf I_\sigma(\x)$
\begin{equation}
\label{eq:Frobenius}	
\rho(\x)=\sqrt{\sum_{i=1}^3\Big(|\partial_x G_\sigma\ast I_i(\x)|^2+|\partial_y G_\sigma\ast I_i(\x)|^2\Big)}.
\end{equation}
For a gray level image $I:\Omega\to\bR$, the edge saliency map $\rho$ can be simply computed  by the norm of the  Euclidean gradient  of the image $I$ such that 
\begin{equation}
\label{eq:EuclideanNorm}
\rho(\x)=\sqrt{|\partial_x G_\sigma\ast I(\x)|^2+|\partial_y G_\sigma\ast I(\x)|^2}.
\end{equation}

\noindent{\textbf{Construction of the Vector Field} $\kg$}.
We use the gradient vector flow method~\cite{xu1998snakes} to compute the vector field  $\kg$  for the construction of  the Randers metric $\cF_\kg$.  This can be  done by minimizing the following functional $\mathcal E_{\rm gvf}$ with respect to a vector field $\vec h=(h_1,h_2)^T:\Omega\to\bR^2$, where  $\mathcal E_{\rm gvf}$ can be expressed as 
\begin{equation}
\label{eq:GVF}
\mathcal E_{\rm gvf}(\vec h)=\epsilon\,\mathcal E_{\rm reg}(\vec h)+\mathcal E_{\rm data}(\vec h),
\end{equation}
where $\epsilon\in\bR^+$ is a  constant and 
\begin{align}
&\mathcal E_{\rm reg}(\vec h)=\int_\Omega\big(\|\nabla h_1(\x)\|^2+\|\nabla h_2(\x)\|^2\big)\,d\x,\\
&\mathcal E_{\rm data}(\vec h)=\int_\Omega\|\nabla\rho(\x)\|\,^2\|\vec h(\x)-\nabla \rho(\x)\|^2\,d\x.
\end{align}
The parameter $\epsilon$  controls the balance between the regularization term $\mathcal E_{\rm reg}$ and the data fidelity term $\mathcal E_{\rm data}$. As discussed in~\cite{xu1998snakes}, the values of $\epsilon$ should depend on the image noise levels  such that a large value of $\epsilon$ is able to suppress the effects from image noise.   We set $\epsilon=0.1$ through all the numerical experiments of this paper. 
The minimization of  the functional $\mathcal E_{\rm gvf}$ can be carried out by solving the  Euler-Lagrange equations of the functional $\mathcal E_{\rm gvf}$ with respect to the components $h_1$ and $h_2$.
The gradient vector flow $\vec h$ is more dense and smooth than  the original gradient filed $\nabla \rho$. In the vicinity of edges, the values of $\|\nabla \rho\|$ are large and  one has the approximation $\vec h\approx\nabla\rho$ due to the effects of  the data fidelity  term $\mathcal E_{\rm data}$, while in the flatten regions where the gradient $\nabla \rho$ nearly vanishes, the vector field $\vec h$ is forced to keep smooth (i.e., slowly-varying),  because within these regions  the minimization of the regularization term $\mathcal E_{\rm reg}$ plays a dominating role for the computation of $\vec h$. More  details for gradient vector flow method can be found from~\cite{xu1998snakes}.

The vector field $\kg$ for the construction of the Randers metric $\cG_\kg$ can be obtained by  normalizing the vector field  $\vec h$ 
\begin{equation}
\label{eq:VectorField}
\kg(\x)=\frac{\vec h(\x)}{\|\vec h(\x)\|},\quad\forall\x\in\Omega.
\end{equation}
The cost functions $\psi_{\rm f}$ and $\psi_{\rm b}$ used in Eq.~\eqref{eq:FinslerianConstruction} for the foreground and background  segmentation application can be expressed  for any $\x\in\Omega$ by
\begin{align}
\label{eq:ObjectSpeed}
&\psi_{\rm f}(\x)=\exp\left(\frac{\alpha_{\rm f}\,\rho(\x)}{\|\rho\|_\infty}\right),\\
\label{eq:ObjectAnisotropy}
&\psi_{\rm b}(\x)=\exp\left(\frac{\alpha_{\rm b}\,\rho(\x)}{\|\rho\|_\infty}\right)\,\psi_{\rm f}(\x),	
\end{align}
where $\alpha_{\rm f}$ and $\alpha_{\rm b}$ are nonnegative constants dominating  how anisotropic and asymmetric the Randers metric $\cG_\kg$ is.  

Once the cost functions $\psi_{\rm f}$ and $\psi_{\rm b}$ have been computed by Eqs.~\eqref{eq:ObjectSpeed} and \eqref{eq:ObjectAnisotropy}, we can construct the tensor filed $\cM_\kg$ and  the vector field $\vec \omega_\kg$ respectively via Eqs.~\eqref{eq:RewrittenTensor} and \eqref{eq:RewrittenVectorFiled}.
Indeed, one has $\psi_{\rm f}(\x)\approx \psi_{\rm b}(\x)\approx 1$ for  the points $\x$  located in the homogeneous region of the image $\mathbf I$ where $\rho(\x)\approx 0$. In this case,  the data-driven Randers metric $\cG_\kg(\x,\cdot)$ expressed in Eq.~\eqref{eq:StaticRandersMetric} approximates to be  an isotropic Riemannian metric. For each point $\x$ around the image edges such that the value of  $\rho(\x)$ is large,  the Randers  metric $\cG_\kg(\x,\cdot)$ will appear to be  strongly anisotropic and asymmetric.

\begin{figure*}[t]
\centering
\includegraphics[width=17cm]{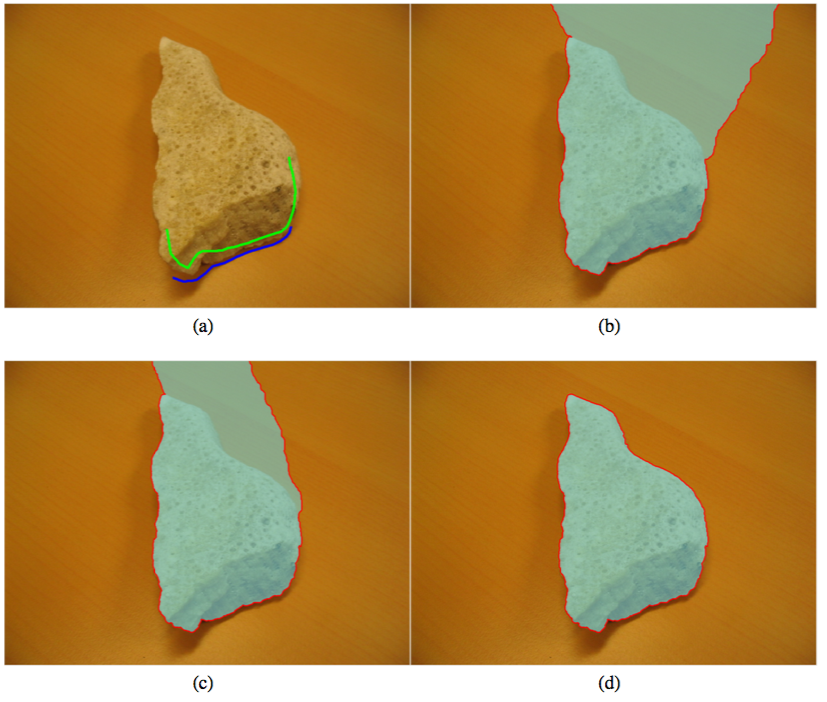}
\caption{Fronts propagation results associated to the Randers metric $\cF_{\kg}$ with different values of $\beta_{\rm d}$. (\textbf{a}) Original image with source points, respectively, placed in the foreground (green brush) and background (blue brush) regions. (\textbf{b})-(\textbf{d}) Fronts propagation for the values of   $\beta_{\rm d}=0,\,5$ and $10$, respectively.}	
\label{fig:VaryDynamic}
\end{figure*}

\subsection{Cost Functions $\psi_{\rm f}$ and $\psi_{\rm b}$ for  Tubularity Segmentation}
\label{subsec:TubularData}
In this section, we take into account  a  feature vector field $\kp:\Omega\to\bR^2$ which extracts local orientation features from the image.  A feature vector $\kp(\x)$ characterizes  the orientation that a tubular structure should have at a point $\x$ belonging to this structure. The feature vector field $\kp$ can be estimated by the steerable filters~\cite{jacob2004design,law2008three,frangi1998multiscale,freeman1991design}. 

Based on the cost functions $\psi_{\rm f}$ and $\psi_{\rm b}$ in Eqs.~\eqref{eq:VectorField} and \eqref{eq:ObjectSpeed}, we are able to  build a positive symmetric definite tensor field $\cM_\kg$ for the Randers metric $\cG_\kg$. For each point $\x$ inside the tubular structures where the gray levels vary slowly, the value of the gradient norm $\rho(\x)$ nearly  vanishes, leading to $\psi_{\rm b}(\x)\approx\psi_{\rm f}(\x)\approx 1$. In this case,  the Randers  metric  $\cG_{\kg}(\x,\fu)$ is  approximately independent  of the directions $\fu$ (see Section~\ref{subsec:ObjectData}).  However, with respect to  tubular structure segmentation application,  we expect that inside the tubular structure the fast marching fronts travel faster along the directions $\kp(\cdot)$ or $-\kp(\cdot)$ than  along the directions $\kp^\perp(\cdot)$ or $-\kp^\perp(\cdot)$ in order to reduce the risk of  the front leakages.  
For this purpose, we make use of  a new tensor field $\tilde\cM_\kg:\Omega\to S_2^+$ based on  $\cM_\kg$ in Eq.~\eqref{eq:RewrittenTensor} such that
\begin{equation*}
\tilde\cM_{\kg}(\x)=\cM_\kg(\x)+\mu\,\kp^\perp(\x)\otimes\kp^\perp(\x),\quad\forall\x\in\Omega,
\end{equation*}
where $\mu\in\bR^+$ is a constant. It  is relevant to the anisotropy property of the tensor field $\tilde\cM_{\kg}$. In the numerical experiments of this paper, we set $\mu=\|\psi_{\rm f}\|^2_\infty$. 

In practice, we observe that inside the tubular structures, the feature vector field $\kp$ can be well approximated by the vector field $\kg^\perp$ which is the orthogonal vector field of $\kg$ derived from Eqs.~\eqref{eq:GVF} and \eqref{eq:VectorField}. In order to reduce the  computation time, we  construct the tensor field $\tilde\cM_\kg$ by the vector field $\kg$ such that
\begin{equation}
\label{eq:EhancedTensorField}
\tilde\cM_{\kg}(\x)=\cM_\kg(\x)+\mu\,\kg(\x)\otimes\kg(\x).
\end{equation}
In this case, we obtain  a new Randers metric $\tilde\cG_\kg$ for the application of  tubularity segmentation, which  depends on the tensor field $\tilde\cM_{\kg}$ and can be formulated as
\begin{equation}
\label{eq:TubularMetric}
\tilde\cG_\kg(\x,\fu):=\sqrt{\langle\fu,\tilde\cM_\kg(\x)\,\fu\rangle}-	\langle\vec\omega_\kg(\x),\fu\rangle.
\end{equation}
Note that we build $\tilde\cG_\kg$ by using  the same  vector field $\vec\omega_\kg$ with the Randers metric $\cG_\kg$. Based on the new  Randers metric $\tilde\cG_\kg$, inside the tubular structures the fast marching fronts will travel fast along the directions $\kp(\cdot)$ or $-\kp(\cdot)$, but slow when the fronts arrive at  the boundaries and slower when the fronts tend to pass through them.

\subsection{Computing the Potential}
\label{subsec:UpdatePotential}
We present the computation methods  for the potential function $\kc$ used by  the data-driven Randers metric that is formulated in Eq.~\eqref{eq:FinslerComponents}.  Basically, the potential function $\kc$ should have small values in the homogeneous regions and large values in the vicinity of the image edges.

\subsubsection{Foreground and Background Segmentation}
\label{subsubsec:FBS} 
For foreground and background segmentation application, the potential function $\kc_{\rm FB}:=\kc$ has a form of  
\begin{equation}
\label{eq:FBPotential}	
\kc_{\rm FB}(\x)=\exp\left(\frac{\beta_{\rm s}\,\rho(\x)}{\|\rho\|_\infty}\right)\,\kc_{\rm dyn}(\x), 
\end{equation} 
where $\beta_{\rm s}$ is a positive constant and $\rho$ is the edge saliency  map defined in Eqs.~\eqref{eq:Frobenius} or \eqref{eq:EuclideanNorm}. 
The term $\exp(\beta_{\rm s}\,\rho(\x))$ which depends only on the edge saliency map $\rho$ will keep invariant during the fast marching fronts propagation. The dynamic potential  function $\kc_{\rm dyn}$ relies on the positions of the  fronts. It will be updated in the course of the geodesic distances computation in terms of some consistency measure of  image features~\cite{bai2009geodesic}. 
Basically, the values of the dynamic potential $\kc_{\rm dyn}$ should be small in the homogeneous regions. We use  a feature map $\kf:\Omega\to\bR^n$ with $n$ the dimensions of the feature vector to establish the dynamic potential  $\kc_{\rm dyn}$. The feature map  $\kf$ can be the image color vector ($n=3$), the image gray level ($n=1$),  or the scalar probability map ($n=1$)~as used in~\cite{bai2009geodesic}.

Recall that in each fast marching distance update iteration, the latest \emph{Accepted} point $\x_{\rm min}$ is chosen by searching for a \emph{Trial} point with  minimal distance value  $\cU_\ks$ ($\ks$ is the set of the source points), i.e., 
\begin{equation}
\label{eq:LatestTrialPoint}
\x_{\rm min}:=	\underset{\x:b(\x)=\emph{Trial}}{\rm{arg\,min}}~\cU_\ks(\x).
\end{equation}
Then the value of $\kc_{\rm dyn}(\z)$ for each point $\z\in\bZ^2\backslash\ks$ such that $\x_{\rm min}\in\Lambda(\z)$ and $b(\z)\neq$\emph{Accepted} can be updated by evaluating the Euclidean  distance between $\kf(\z)$ and $\kf(\x_{\rm min})$  (see Line~\ref{line:DynamicUpdate} of Algorithm~\ref{alg:VIM}). In other words, one can compute  the dynamic potential $\kc_{\rm dyn}$ in each fast marching  update iteration by
\begin{equation}
\label{eq:DynamicSpeed}
\kc_{\rm dyn}(\z)=\exp(\,\beta_{\rm d}\,\|\kf(\z)-\kf(\x_{\rm min})\|\,)	
\end{equation}
for all grid points $\z\in\bZ^2\backslash\ks$ such that $\x_{\rm min}\in\Lambda(\z)$ and $b(\z)\neq$\emph{Accepted}, where $\beta_{\rm d}$ is a positive constant. Note that  we initialize the dynamic potential $\kc_{\rm dyn}$ by
\begin{equation*}
\kc_{\rm dyn}(\x)=1,\quad \forall \x\in\ks.
\end{equation*}

Based on the potential $\kc_{\rm FB}$ in Eq.~\eqref{eq:FBPotential} and the Randers  metric $\cG_\kg$ (see Section~\ref{subsec:ObjectData}), the data-driven Randers metric $\cF_\kg$ for the  foreground and background segmentation application can be expressed for any point $\x\in\Omega$ and any vector $\fu\in\bR^2$ by
\begin{equation}
\label{eq:FinalFBRanders}
\cF_\kg(\x,\fu)=\kc_{\rm FB}(\x)\,\cG_{\kg}(\x,\fu).
\end{equation}

Finally, we show the effects of the dynamic potential $\kc_{\rm dyn}$ in Eq.~\eqref{eq:DynamicSpeed} in  the foreground and background segmentation application.  The fronts propagation results are shown in Fig.~\eqref{fig:VaryDynamic} with respect to the Randers metric $\cF_{\kg}$. In this experiment, we choose different values for the parameter  $\beta_{\rm d}$ that is used by the dynamic potential $\kc_{\rm dyn}$ and a fixed parameter $\beta_{\rm s}=10$ to compute the edge-based potential $\kc_{\rm FB}$. The values of $\alpha_{\rm f}$ and $\alpha_{\rm b}$ are set to be $2$ and $3$, respectively. 
Indeed, one can point out  that the dynamic potential   is able to help the fronts propagation scheme to avoid leakages.

\subsubsection{Tubularity Segmentation} 
We assume that the gray levels inside the tubular structures are stronger than outside.  For  tubularity segmentation, we consider a potential function $\kc_{\rm T}:=\kc$ involving a static term and a dynamic term $\tilde\kc_{\rm dyn}$ such that
\begin{equation}
\label{eq:TubePotential}	
\kc_{\rm T}(\x)=\exp\big(\beta_{\rm s}(\|\zeta\|_\infty-\zeta(\x))\big)\,\tilde\kc_{\rm dyn}(\x), 
\end{equation} 
where  $\zeta:\Omega\to[0,1]$ is a feature map that  characterizes the tubularity appearance, i.e.,  the value of $\zeta(\x)$ indicates  the probability of a point $\x$  belonging to a tubular structure.  In practice,  the map $\zeta$ can be specified as the image gray levels or as  a normalized vesselness map derived from a tubularity detector such as~\cite{frangi1998multiscale,franken2009crossing,law2008three}.

The dynamic potential  $\tilde\kc_{\rm dyn}$ is estimated in the course of the fast marching fronts propagation.  The computation scheme of $\tilde\kc_{\rm dyn}$ is  almost identical to  the dynamic potential $\kc_{\rm dyn}$  presented in Section~\ref{subsubsec:FBS}, except that the dynamic potential $\tilde\kc_{\rm dyn}$ for tubularity  segmentation is dependent on the feature map $\zeta$.  

We also initialize the dynamic potential by $\tilde\kc_{\rm dyn}(\x)=1,~\forall\x\in\ks$. 
In each geodesic distance update iteration, we suppose   $\x_{\rm min}$ be the latest \emph{Accepted} point. For each grid point $\z\in\bZ^2\backslash\ks$ such that $\x_{\rm min}\in\Lambda(\z)$ and $b(\z)\neq$\,\emph{Accepted}, we update the dynamic potential $\tilde\kc_{\rm dyn}$ by
\begin{align}
\label{eq:VesselnessConsistency}
\tilde\kc_{\rm dyn}(\z)=\exp(\,\beta_{\rm d}\,|\min\{\zeta(\z)-\zeta(\x_{\rm min}),0\}|\,).
\end{align}
The reason for the use of \eqref{eq:VesselnessConsistency} is that  if  the latest \emph{Accepted} point $\x_{\rm min}$ belongs to a tubular structure, then the inequality  $\zeta(\z)>\zeta(\x_{\rm min})$ means that the neighbour point $\z$ also belongs to this structure. Therefore,  we assign a small value to $\tilde\kc_{\rm dyn}$.
 In this paper, we set the map $\zeta$ as the normalized image gray levels, i.e., $\zeta(\x)\in[0,1],~\forall\x\in\Omega$. 

The data-driven Randers metric $\tilde\cF_\kg$ for tubularity segmentation can be formulated by
\begin{equation}
\label{eq:FinalTubeRanders}
\tilde\cF_\kg(\x,\fu)=\kc_{\rm T}(\x)\,\tilde\cG_\kg(\x,\fu), \quad \forall\x\in\Omega,\,\forall\fu\in\bR^2,
\end{equation}
where the potential $\kc_{\rm T}$ and the metric $\tilde\cG_\kg$ are expressed in Eqs.~\eqref{eq:TubePotential}  and \eqref{eq:TubularMetric}, respectively.

\begin{figure*}[t]
\setcounter{subfigure}{0}
\centering
\includegraphics[height=5.5cm]{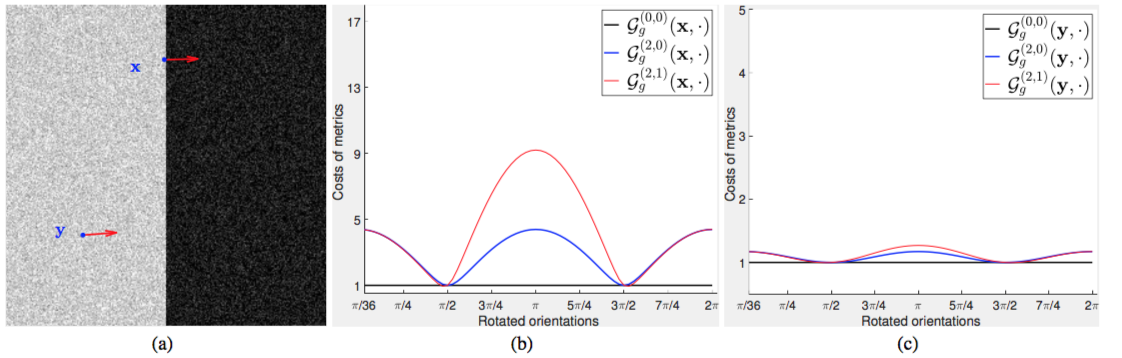}
\caption{\textbf{a} shows a synthetic image. The blue dots indicate two sampled points. The arrows indicate the directions $\kg(\x)$ and $\kg(\y)$. \textbf{b} and \textbf{c} Plots of the cost values of $\cG^{(0,0)}_\kg$, $\cG^{(2,0)}_\kg$ and $\cG^{(2,1)}_\kg$ at points $\x$ and $\y$ along different directions.}
\label{fig:EdgePlot}
\end{figure*}

\begin{figure*}[t]
\setcounter{subfigure}{0}
\centering
\includegraphics[height=5.5cm]{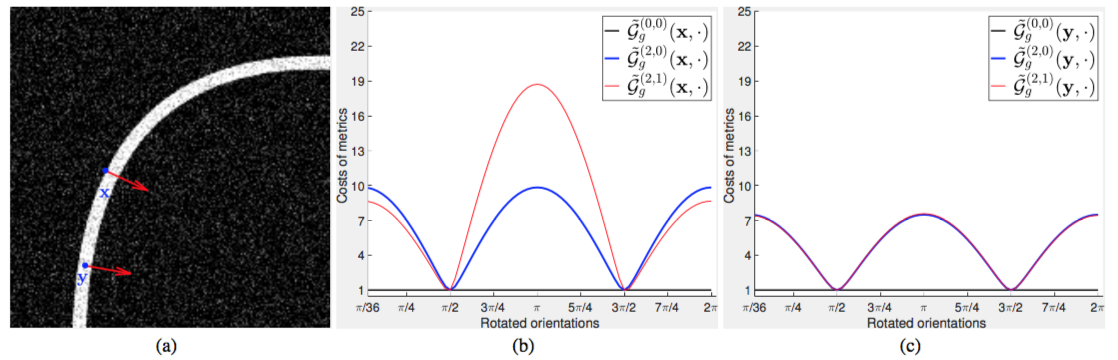}
\caption{\textbf{a} A synthetic image. The blue dots indicate two sampled points. The arrows indicate the  directions of  $\kg(\x)$ and $\kg(\y)$. \textbf{b} and \textbf{c} Plots of the cost values of $\tilde\cG^{(0,0)}_\kg$, $\tilde\cG^{(2,0)}_\kg$ and $\tilde\cG^{(2,1)}_\kg$ at points $\x$ and $\y$ along different directions.}
\label{fig:TubularPlot}
\end{figure*}

\section{Experimental Results}
\label{sec:Experiment}
\subsection{Implementation Details}

\begin{figure*}[t]
\centering
\includegraphics[width=17cm]{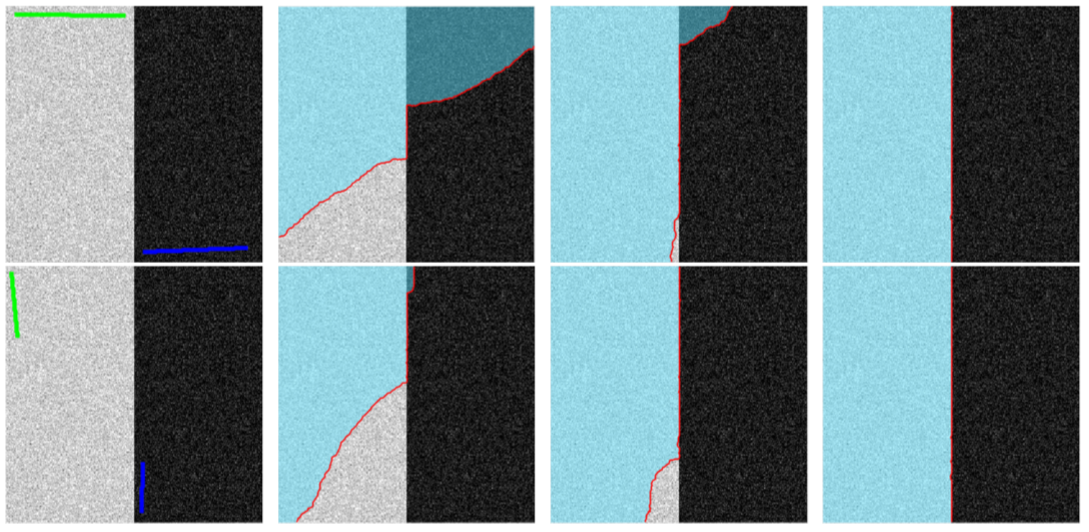}
\caption{Image segmentation via different geodesic metrics on a synthetic image. \textbf{Column 1} shows the initializations, where the green and blue brushes indicating the seeds in different regions. \textbf{Columns 2-4} show the segmentation results by the fronts propagation associated to  the isotropic Riemannian metric $\cF^{(0,0)}_{\kg}$, the anisotropic Riemannian metric $\cF^{(2,0)}_\kg$ and the Randers metric $\cF_{\kg}^{(2,3)}$, respectively. The blue curves are the segmented  boundaries of the Voronoi regions.}
\label{fig:SynVoronoi}
\end{figure*}

\noindent{\textbf{Parameter Setting.}}
The anisotropy and asymmetry of the Randers metrics $\cG_\kg$  and $\tilde\cG_\kg$ are determined by  the parameters $\alpha_{\rm f}$ and $\alpha_{\rm b}$ (see Eqs.~\eqref{eq:ObjectSpeed} and \eqref{eq:ObjectAnisotropy}).  We respectively  denote  by $\cG_\kg^{\vec\alpha}$ and $\tilde\cG_\kg^{\vec \alpha}$ the data-driven Randers metrics $\cG_{\kg}$ and $\tilde\cG_\kg$ with a pair of parameters $\vec\alpha=(\alpha_{\rm f},\alpha_{\rm b})$. In this case, the corresponding  Randers metrics $\cF_{\kg}$ in Eq.~\eqref{eq:FinalFBRanders}  and $\tilde\cF_\kg$ in~\eqref{eq:FinalTubeRanders} can be noted  by $\cF^{\vec\alpha}_{\kg}$ and $\tilde\cF^{\vec\alpha}_\kg$, respectively.  The potential functions $\kc_{\rm FB}$ and $\kc_{\rm T}$ rely on two parameters $\beta_{\rm s}$ and $\beta_{\rm d}$. We fix $\beta_{\rm d}=10$ through all the experiments,  except in Fig.~\ref{fig:SynVoronoi} for which we set $\beta_{\rm d}=5$. The values of $\beta_{\rm s}$ are individually set  for each experiment.

Note that the parameter $\vec\alpha=(0,0)$ indicates that the geodesic metrics $\cG^{(0,0)}_\kg$ and $\tilde\cG^{(0,0)}_\kg$ are isotropic with respect to the second argument. Furthermore, when  $\vec\alpha=(a,0)$ with $a\in\bR^+$, the metrics $\cG_\kg^{(a,0)}$ and $\tilde\cG_\kg^{(a,0)}$ get to be the anisotropic Riemannian cases\footnote{Note that metrics $\cG_\kg^{\vec \alpha}$ and $\tilde\cG_\kg^{\vec \alpha}$ have the identical  anisotropy and asymmetry properties to $\cF_\kg^{\vec\alpha}$ and $\tilde\cF_\kg^{\vec\alpha}$, respectively.}. 

\noindent{\textbf{Image Segmentation Schemes.}}
The interactive foreground and background  segmentation task can be converted to the problem of  identifying the Voronoi index map or Voronoi regions  in terms of geodesic distance~\cite{bai2009geodesic,arbelaez2004energy}.
  Let $\ks_1$ and $\ks_2$ be the sets of source points which are  respectively located at the foreground and background regions. The Voronoi regions $\cV_1$ and $\cV_2$, indicating foreground and background regions respectively,  can be determined by the Voronoi index map $\cL$ through Eq.~\eqref{eq:VR} such that 
\begin{equation*}
\cV_i:=\{\x\in\Omega;\cL(\x)=i\},\quad i=1,\,2.  	
\end{equation*}

With respect to the foreground and background segmentation, the computation complexity for the geodesic distance   computation is consistent to the used  fast marching method~\cite{mirebeau2014efficient}, which is bounded by $\mathcal O(N\ln\kappa(\cF_\kg)+N\ln N)$ with $N$ the number of the grid  points in $\bZ^2$. 

For tubularity segmentation,  all the user-provided seeds are supposed to be placed inside the targeted structures. We make use of the  $T_h$-level set of the geodesic distance map $\cU_\ks$ as the boundaries of the targeted tubular structures.  We denote by $N_{\rm accepted}$  the number of  the grid points in $\bZ^2$ tagged as \emph{Accepted}, i.e., 
\begin{equation*}
N_{\rm accepted}=\#\{\x\in\bZ^2;\,b(\x)=\emph{Accepted}\}.	
\end{equation*}
In order to reduce the computation time of the fast marching method, we terminate the fronts propagation once the number $N_{\rm accepted}$ of the grid points tagged as \emph{Accepted} reaches the given thresholding value $N_{\rm  th}$.  
In this case, the tubular structures can be  recovered by the regions which are comprised of all the \emph{Accepted} points.  Let $N_{\rm trial}$ be the number of \emph{Trial} grid points when $N_{\rm accepted}$ points have been tagged as \emph{Accepted} and let $N_{\rm used}=N_{\rm accepted}+N_{\rm trial}$ be the total number of grid points for which the distance values have been updated. The computation complexity for this application is bounded by $\mathcal O(N_{\rm used}\ln\kappa(\tilde\cF_\kg)+N_{\rm used}\ln N_{\rm used})$, since only $N_{\rm used}$ grid points are visited by the fast marching fronts.

\begin{figure*}[p]
\centering
\includegraphics[width=17cm]{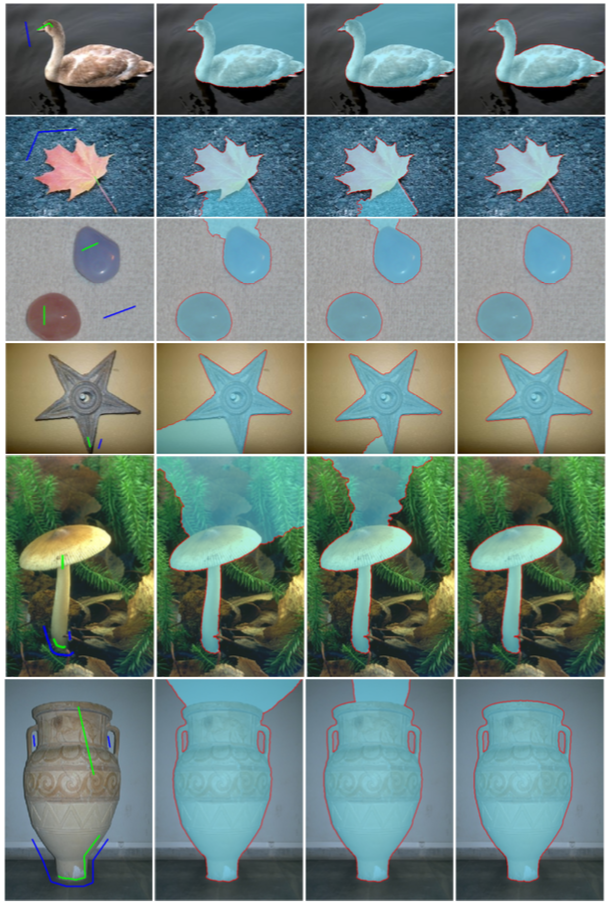}
\caption{Image segmentation via different geodesic metrics on real images. \textbf{Column 1} shows the initializations, where the green and blue brushes are the seeds indicate the background and foreground regions. \textbf{Columns 2-4} show the segmentation results by the metrics $\cF^{(0,0)}_\kg$,  $\cF^{(2,0)}_\kg$ and  $\cF_\kg^{(2,3)}$, respectively. The red  curves are the segmented boundaries of the respective  Voronoi regions.}
\label{fig:Voronoi}
\end{figure*}

\begin{figure*}[t]
\centering
\includegraphics[width=17cm]{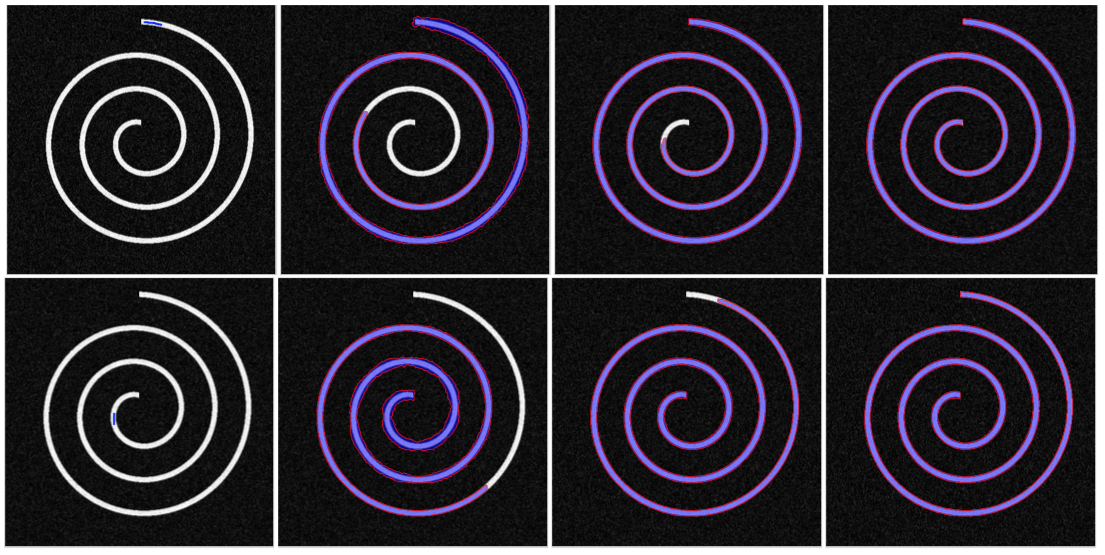}
\caption{Tubularity segmentation results via different geodesic metrics on a Spiral. \textbf{Column 1} shows the initializations with  the blue brushes indicating the seeds. \textbf{Columns 2-4} show the segmentation results through the geodesic metrics $\tilde\cF^{(0,0)}_\kg$,  $\tilde\cF^{(2,0)}_\kg$ and $\tilde\cF_\kg^{(2,3)}$, respectively.}
\label{fig:Spiral}
\end{figure*}

\begin{figure*}[t]
\centering
\includegraphics[width=17cm]{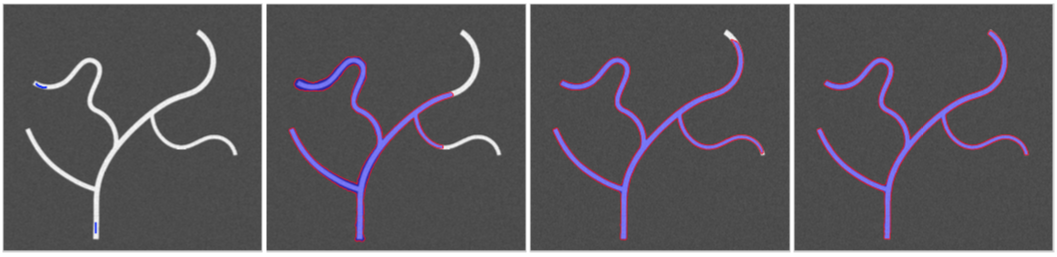}
\caption{Tubularity segmentation results via different geodesic metrics on a tubular tree. \textbf{Column 1} shows the initializations with  the blue brushes indicating the seeds. \textbf{Columns 2-4} show the segmentation results through the geodesic metrics $\tilde\cF^{(0,0)}_\kg$,  $\tilde\cF^{(2,0)}_\kg$ and $\tilde\cF_\kg^{(2,3)}$, respectively.}
\label{fig:Tree}
\end{figure*}

\subsection{Advantages of Using  Anisotropy and Asymmetry Enhancements}
Let us consider a synthetic  image as shown in Fig.~\ref{fig:EdgePlot}a with two sampled points $\x$ and $\y$ indicated by blue dots. The arrows respectively indicate the directions of $\kg(\x)$ and $\kg(\y)$, where $\x$ is near the edges and $\y$ is located inside the homogeneous region.  In Fig.~\ref{fig:EdgePlot}b, we plot the cost values of the metrics $\cG^{(0,0)}_\kg(\x,\fu_j)$,  $\cG_\kg^{(2,0)}(\x,\fu_j)$ and  $\cG^{(2,1)}_\kg(\x,\fu_j)$,  along different directions $\vec u_j\in\bR^2$. The directions $\fu_j$ are obtained by rotation such that 
\begin{equation}
\label{eq:RotatedAnkles}
\vec u_j=M(j\,\theta_{\rm s})\,\kg(\x),\quad j=1,2,...,72,
\end{equation}
where $\theta_{\rm s}=\pi/36$ is the angle resolution and $M(j\,\theta_{\rm s})$ is a  rotation matrix with angle $j\,\theta_{\rm s}$ in a countclockwise order. In Fig.~\ref{fig:EdgePlot}c, we plot the cost values for the metrics  $\cG^{(0,0)}_\kg(\y,\vec v_j)$, $\cG^{(2,0)}_\kg(\y,\fv_j)$ and $\cG^{(2,1)}_\kg(\y,\vec v_j)$ with  
\begin{equation*}
\vec v_j=M(j\,\theta_{\rm s})\kg(\y).
\end{equation*}
In Fig.~\ref{fig:EdgePlot}b, we can see that all of the three metrics  get low values around the  directions $M(\pi/2)\,\kg(\x)$ and $M(3\pi/2)\,\kg(\x)$, which are orthogonal to the direction $\kg(\x)$. However, around the direction $-\kg(\x)$, the Randers metric $\cG^{(2,1)}_\kg$ attains much larger values than the Riemannian cases  $\cG_\kg^{(0,0)}$ and $\cG_\kg^{(2,0)}$. Such an asymmetric property is able to reduce the risk of   front leakages. 

In Fig.~\ref{fig:TubularPlot}, we plot the cost values for the metrics $\tilde\cG^{(0,0)}_\kg$, $\tilde\cG^{(2,0)}_\kg$ and $\tilde\cG^{(2,1)}_\kg$ at the points $\x$ and $\y$ along different rotated directions. In Fig.~\ref{fig:TubularPlot}b, the cost values (indicated by the red curve) of  the Randers metric  $\tilde\cG_\kg^{(2,1)}$ at  the point $\x$  near the edges and outside the tubular structure  show strongly asymmetric property. In Fig.~\ref{fig:TubularPlot}c, we can see that along all of the rotated directions, the  cost values for both the Randers metric $\tilde\cG_\kg^{(2,1)}$ are equivalent to  the anisotropic Riemannian metric $\tilde\cG_\kg^{(2,0)}$. This anisotropic property is able to force the fast marching fronts travel faster along the tubularity orientations which are approximated by the vector field  $\kg^\perp$. 

\subsection{Experiments on  Synthetic and Real Images}
In Fig.~\ref{fig:SynVoronoi}, we show the fronts propagation results on a synthetic image. In the first column of Fig.~\ref{fig:SynVoronoi}, we initialize the sets of the  source points  in different locations, which are indicated by green and blue brushes. The columns $2$ to $4$ of  Fig.~\ref{fig:SynVoronoi} are the segmentation results from the isotropic Riemannian metric $\cF_\kg^{(0,0)}$, the anisotropic Riemannian  metric $\cF^{(2,0)}_\kg$ and the Randers metric $\cF^{(2,3)}_\kg$, respectively. For the purpose of  fair comparisons, the three metrics used in this experiment share the same potential function $\kc$ defined in Eq.~\eqref{eq:FBPotential}. One can point out that the results from the metrics $\cF^{(0,0)}_\kg$ and  $\cF^{(2,0)}_\kg$ suffer from the leaking problem,  while the final boundaries (red curves) associated the proposed Randers metric $\cF^{(2,3)}_\kg$ are able to catch the expected edges thanks to the asymmetric enhancement.  In this experiment, we choose $\beta_{\rm d}=5$.

In Fig.~\ref{fig:Voronoi}, we compare the interactive image segmentation results  via different geodesic metrics on real images which are obtained from the Weizmann dataset~\cite{alpert2012image} and the Grabcut dataset~\cite{rother2004grabcut}. The final segmentation results  are derived from  the boundaries of the corresponding Voronoi index maps. In column 1, we show the initial images with seeds indicating by green  and blue  brushes respectively inside the foreground and background regions. In columns $2$ to $4$ of  Fig.~\ref{fig:Voronoi}, we demonstrate the segmentation results obtained  via the isotropic Riemannian  metric $\cF_\kg^{(0,0)}$, the anisotropic Riemannian metric $\cF_\kg^{(2,0)}$ and the Randers Metric $\cF_\kg^{(2,3)}$. For the results from the isotropic and anisotropic Riemannian metrics (shown in columns $2$ and $3$), the final contours leak into the background regions. In contrast, the segmentation contours associated to the Randers metric $\cF^{(2,3)}$ are able to follow the desired object boundaries.

In Fig.~\ref{fig:Spiral},  we compare the tubularity segmentation results  on a Spiral respectively using the isotropic Riemannian metric $\tilde\cF^{(0,0)}_\kg$, the anisotropic Riemannian metric $\tilde\cF^{(2,0)}_\kg$ and the Randers metric $\tilde\cF^{(2,3)}_\kg$. The tubularity boundaries (red curves) are computed through the level set lines of the respective  geodesic distance maps with an identical thresholding value of $N_{\rm th}$. The  shadow regions indicate the segmented regions involving all the points tagged as \emph{Accepted}. In the first column of Fig.~\ref{fig:Spiral}, for each row the respective source points  are placed at the  end of the Spiral. One can point out that  the final segmentation contours corresponding to the isotropic Riemannian metric $\tilde\cF_\kg^{(0,0)}$ (shown in column $2$)  and the anisotropic Riemannian metric $\tilde\cF^{(2,0)}_\kg$ (shown in column $3$) leak into the background before the whole tubular structure has been covered by the fast marching fronts. Indeed, using the anisotropy enhancement can improve the segmentation  results such that the leakages for the contours from the anisotropic Riemannian metric $\tilde\cF_\kg^{(2,0)}$ only occur at the locations far from the seeds. Finally, the segmentation contours shown in column $4$  resulted by the Randers metric $\cF^{(2,3)}$ with both anisotropic and asymmetric enhancements are able to delineate the desired  tubularity boundaries.

In Fig.~\ref{fig:Tree}, we perform the fast marching  fronts propagation on a tubular tree structure. We again observe the leaking problem that occurs in the segmentation  results derived from the isotropic Riemannian metric $\tilde\cF^{(0,0)}_\kg$ and the anisotropic Riemannian metric $\tilde\cF_\kg^{(2,0)}$, which are shown in columns 2 and 3 respectively.  While in column $4$, the fronts are able to pass through  the whole tubular tree structure before they  leak  into the background.

\section{Conclusion}
\label{sec:Conclusion}
In this paper, we extend the fronts propagation framework from the  Riemannian case to  a general Finsler case with applications to image segmentation. The Finsler metric with a Randers form allows us to  take into account the asymmetric and anisotropic image features in order to reduce the risk of the leaking problem during  the fronts propagation.  We presented a method for  the construction of the Finsler  metric with a Randers form using  a vector field derived from the image edges. This metric can also integrate with a feature coherence penalization term updated in the course of  the fast marching fronts propagation.    We applied the fronts propagation model associated to the proposed Randers metrics to  foreground and background segmentation and tubularity segmentation. Experimental results show that the proposed model indeed produces promising results.

\section*{Acknowledgment}
The authors would like to thank all the anonymous reviewers for their detailed remarks that helped us improve the presentation of this paper. The authors  thank Dr. Jean-Marie Mirebeau from Universit\'e Paris-Sud for his fruitful discussion and creative suggestions. The first author also  thanks Dr. Gabriel Peyr\'e from ENS Paris for his financial support. This work was partially supported by the European Research Council (ERC project SIGMA-Vision).

\bibliographystyle{spmpsci}      % mathematics and physical sciences
\bibliography{FrontPropagation}   % name your BibTeX data base

\begin{thebibliography}{10}
\providecommand{\url}[1]{{#1}}
\providecommand{\urlprefix}{URL }
\expandafter\ifx\csname urlstyle\endcsname\relax
  \providecommand{\doi}[1]{DOI~\discretionary{}{}{}#1}\else
  \providecommand{\doi}{DOI~\discretionary{}{}{}\begingroup
  \urlstyle{rm}\Url}\fi

\bibitem{adalsteinsson1995fast}
Adalsteinsson, D., Sethian, J.A.: A fast level set method for propagating
  interfaces.
\newblock Journal of Computational Physics \textbf{118}(2), 269--277 (1995)

\bibitem{alpert2012image}
Alpert, S., Galun, M., Brandt, A., Basri, R.: Image segmentation by
  probabilistic bottom-up aggregation and cue integration.
\newblock IEEE Trans. on Pattern Analysis and Machine Intelligence
  \textbf{34}(2), 315--327 (2012)

\bibitem{arbelaez2008constrained}
Arbel{\'a}ez, P., Cohen, L.: Constrained image segmentation from hierarchical
  boundaries.
\newblock In: Proceedings of CVPR, pp. 1--8 (2008)

\bibitem{arbelaez2004energy}
Arbel{\'a}ez, P.A., Cohen, L.D.: Energy partitions and image segmentation.
\newblock Journal of Mathematical Imaging and Vision \textbf{20}(1), 43--57
  (2004)

\bibitem{bai2007geodesic}
Bai, X., Sapiro, G.: A geodesic framework for fast interactive image and video
  segmentation and matting.
\newblock In: Proceedings of ICCV 2007, pp. 1--8 (2007)

\bibitem{bai2009geodesic}
Bai, X., Sapiro, G.: Geodesic matting: A framework for fast interactive image
  and video segmentation and matting.
\newblock International Journal of Computer Vision \textbf{82}(2), 113--132
  (2009)

\bibitem{benmansour2009fast}
Benmansour, F., Cohen, L.D.: Fast object segmentation by growing minimal paths
  from a single point on 2d or 3d images.
\newblock Journal of Mathematical Imaging and Vision \textbf{33}(2), 209--221
  (2009)

\bibitem{bornemann2006finite}
Bornemann, F., Rasch, C.: {Finite-element discretization of static
  Hamilton-Jacobi equations based on a local variational principle}.
\newblock Computing and Visualization in Science \textbf{9}(2), 57--69 (2006)

\bibitem{cardinal2006intravascular}
Cardinal, M.H., Meunier, J., et~al.: Intravascular ultrasound image
  segmentation: a three-dimensional fast-marching method based on gray level
  distributions.
\newblock IEEE Trans. on Medical Imaging \textbf{25}(5), 590--601 (2006)

\bibitem{caselles1993geometric}
Caselles, V., Catt{\'e}, F., Coll, T., Dibos, F.: A geometric model for active
  contours in image processing.
\newblock Numerische Mathematik \textbf{66}(1), 1--31 (1993)

\bibitem{caselles1997geodesic}
Caselles, V., Kimmel, R., Sapiro, G.: Geodesic active contours.
\newblock International Journal of Computer Vision \textbf{22}(1), 61--79
  (1997)

\bibitem{chen2016vessel}
Chen, D., Cohen, L.D.: Vessel tree segmentation via front propagation and
  dynamic anisotropic riemannian metric.
\newblock In: Proceedings of ISBI, pp. 1131--1134 (2016)

\bibitem{chen2016finsler}
Chen, D., Mirebeau, J.M., Cohen, L.D.: Finsler geodesic evolution model for
  region-based active contours.
\newblock In: Proceedings of BMVC (2016)

\bibitem{chen2017global}
Chen, D., Mirebeau, J.M., Cohen, L.D.: {Global minimum for a Finsler elastica
  minimal path approach}.
\newblock International Journal of Computer Vision \textbf{122}(3), 458--483
  (2017)

\bibitem{cohen2001multiple}
Cohen, L.: Multiple contour finding and perceptual grouping using minimal
  paths.
\newblock Journal of Mathematical Imaging and Vision \textbf{14}(3), 225--236
  (2001)

\bibitem{cohen1991active}
Cohen, L.D.: On active contour models and balloons.
\newblock CVGIP: Image Understanding \textbf{53}(2), 211--218 (1991)

\bibitem{cohen2007segmentation}
Cohen, L.D., Deschamps, T.: {Segmentation of 3D tubular objects with adaptive
  front propagation and minimal tree extraction for 3D medical imaging}.
\newblock Computer Methods in Biomechanics and Biomedical Engineering
  \textbf{10}(4), 289--305 (2007)

\bibitem{cohen1997global}
Cohen, L.D., Kimmel, R.: {Global minimum for active contour models: A minimal
  path approach}.
\newblock International Journal of Computer Vision \textbf{24}(1), 57--78
  (1997)

\bibitem{criminisi2008geos}
Criminisi, A., Sharp, T., Blake, A.: {Geos: Geodesic image segmentation}.
\newblock In: Proceedings of ECCV, pp. 99--112 (2008)

\bibitem{dijkstra1959note}
Dijkstra, E.W.: A note on two problems in connexion with graphs.
\newblock Numerische Mathematik \textbf{1}(1), 269--271 (1959)

\bibitem{kass1988snakes}
Kass, M., Witkin, A., Terzopoulos, D.: {Snakes: Active contour models}.
\newblock International Journal of Computer Vision \textbf{1}(4), 321--331
  (1988)

\bibitem{kimmel2003regularized}
Kimmel, R., Bruckstein, A.M.: Regularized laplacian zero crossings as optimal
  edge integrators.
\newblock International Journal of Computer Vision \textbf{53}(3), 225--243
  (2003)

\bibitem{li2010distance}
Li, C., Xu, C., Gui, C., Fox, M.D.: Distance regularized level set evolution
  and its application to image segmentation.
\newblock IEEE Trans. on Image Processing \textbf{19}(12), 3243--3254 (2010)

\bibitem{li2007local}
Li, H., Yezzi, A.: {Local or global minima: Flexible dual-front active
  contours}.
\newblock IEEE Trans. on Pattern Analysis and Machine Intelligence
  \textbf{29}(1) (2007)

\bibitem{malladi1995shape}
Malladi, R., Sethian, J., Vemuri, B.C.: {Shape modeling with front propagation:
  A level set approach}.
\newblock IEEE Trans. on Pattern Analysis and Machine Intelligence
  \textbf{17}(2), 158--175 (1995)

\bibitem{malladi1998real}
Malladi, R., Sethian, J.A.: A real-time algorithm for medical shape recovery.
\newblock In: Proceeding of ICCV, pp. 304--310 (1998)

\bibitem{melonakos2008finsler}
Melonakos, J., Pichon, E., Angenent, S., Tannenbaum, A.: Finsler active
  contours.
\newblock IEEE Trans. on Pattern Analysis and Machine Intelligence
  \textbf{30}(3), 412--423 (2008)

\bibitem{mirebeau2014anisotropic}
Mirebeau, J.M.: Anisotropic fast-marching on cartesian grids using lattice
  basis reduction.
\newblock SIAM Journal on Numerical Analysis \textbf{52}(4), 1573--1599 (2014)

\bibitem{mirebeau2014efficient}
Mirebeau, J.M.: {Efficient fast marching with Finsler metrics}.
\newblock Numerische Mathematik \textbf{126}(3), 515--557 (2014)

\bibitem{mirebeau2017anisotropic}
Mirebeau, J.M.: Anisotropic fast-marching on cartesian grids using voronoi's
  first reduction of quadratic forms.
\newblock Preprint  (2017)

\bibitem{osher1988fronts}
Osher, S., Sethian, J.A.: {Fronts propagating with curvature-dependent speed:
  algorithms based on Hamilton-Jacobi formulations}.
\newblock JCP \textbf{79}(1), 12--49 (1988)

\bibitem{price2010geodesic}
Price, B.L., Morse, B., Cohen, S.: Geodesic graph cut for interactive image
  segmentation.
\newblock In: Proceedings of CVPR, pp. 3161--3168 (2010)

\bibitem{randers1941asymmetrical}
Randers, G.: On an asymmetrical metric in the four-space of general relativity.
\newblock Physical Review \textbf{59}(2), 195 (1941)

\bibitem{rother2004grabcut}
Rother, C., Kolmogorov, V., Blake, A.: {Grabcut: Interactive foreground
  extraction using iterated graph cuts}.
\newblock ACM Trans. on Graphics \textbf{23}(3), 309--314 (2004)

\bibitem{rouy1992viscosity}
Rouy, E., Tourin, A.: A viscosity solutions approach to shape-from-shading.
\newblock SIAM Journal on Numerical Analysis \textbf{29}(3), 867--884 (1992)

\bibitem{sethian1999fast}
Sethian, J.A.: Fast marching methods.
\newblock SIAM Review \textbf{41}(2), 199--235 (1999)

\bibitem{sethian2003ordered}
Sethian, J.A., Vladimirsky, A.: {Ordered upwind methods for static
  Hamilton--Jacobi equations: Theory and algorithms}.
\newblock SIAM Journal on Numerical Analysis \textbf{41}(1), 325--363 (2003)

\bibitem{tsitsiklis1995efficient}
Tsitsiklis, J.N.: Efficient algorithms for globally optimal trajectories.
\newblock IEEE Transactions on Automatic Control \textbf{40}(9), 1528--1538
  (1995)

\bibitem{weber2008parallel}
Weber, O., Devir, Y.S., et~al.: Parallel algorithms for approximation of
  distance maps on parametric surfaces.
\newblock ACM Trans. on Graphics \textbf{27}(4), 104 (2008)

\bibitem{xu1998snakes}
Xu, C., Prince, J.L.: Snakes, shapes, and gradient vector flow.
\newblock IEEE Trans. on Image Processing \textbf{7}(3), 359--369 (1998)

\bibitem{yatziv2006n}
Yatziv, L., Bartesaghi, A., Sapiro, G.: {O (N) implementation of the fast
  marching algorithm}.
\newblock Journal of Computational Physics \textbf{212}(2), 393--399 (2006)

\bibitem{yezzi1997geometric}
Yezzi, A., Kichenassamy, S., Kumar, A., Olver, P., Tannenbaum, A.: A geometric
  snake model for segmentation of medical imagery.
\newblock IEEE Trans. on Medical Imaging \textbf{16}(2), 199--209 (1997)

\end{thebibliography}

%%\begin{thebibliography}{}
%%\bibitem{RefJ}
%%Author, Article title, Journal, Volume, page numbers (year)
%%\bibitem{RefB}
%%Author, Book title, page numbers. Publisher, place (year)
% etc
%%\end{thebibliography}

\end{document}